\documentclass[sn-mathphys-num]{sn-jnl}

\usepackage{graphicx}%
\usepackage{multirow}%
\usepackage{amsmath,amssymb,amsfonts}%
\usepackage{amsthm}%
\usepackage{mathrsfs}%
\usepackage[title]{appendix}%
\usepackage{xcolor}%
\usepackage{textcomp}%
\usepackage{manyfoot}%
\usepackage{booktabs}%
\usepackage{hyperref}
\hypersetup{hypertexnames=false}%
\usepackage{mathtools}
\usepackage{bm}
\usepackage[capitalize]{cleveref}
\usepackage{siunitx}
\usepackage{subcaption}
\usepackage{pgfplots}
\usepackage{jsonparse}
\usepackage{autonum}
\makeatletter
\autonum@generatePatchedReferenceCSL{Cref}
\makeatother
\usepackage[nohypertypes={acronym,main}]{glossaries-prefix}

\let\originalleft\left
\let\originalright\right
\renewcommand{\left}{\mathopen{}\mathclose\bgroup\originalleft}
\renewcommand{\right}{\aftergroup\egroup\originalright}

\newcommand{\grad}{\vec \nabla}
\newcommand{\vel}{\vec u}

\newcommand{\mDDt}[1]{\frac{\mathrm{D}#1}{\mathrm{D} t}}

\newcommand{\pd}[2]{\frac{\partial#1}{\partial #2}}

\newcommand{\transp}{^{\mathbf{T}}}

\renewcommand{\vec}[1]{ \bm{ \mathbf{ #1 } } }
\newcommand{\mat}[1]{ \vec{#1} } %
\newcommand{\matDer}[1]{\pd{#1}{t} + (\vel \cdot \grad) #1}
\newcommand{\symVelGrad}{\frac{1}{2} \left( \big( \grad \vel \big) + \big( \grad \vel \big)\transp \right)}

\newcommand{\trafoCoord}[1]{\Qm\transp #1 \Qm}

\newcommand{\strain}{\mat{E}}

\newcommand{\straint}{\mat{E}^{\Qm}}

\newcommand{\straintI}[1]{E^{\Qm}_{#1}}
\newcommand{\strainI}[1]{E_{#1}}

\newcommand{\Qm}{\mat{\Qsymbol}}

\newcommand{\Qsymbol}{V}
\newcommand{\Lamb}{\mat{\Lambda}}
\newcommand{\diag}{\mat{\mathrm{diag}}}

\newcommand{\eigval}{\lambda}

\newcommand{\logLambI}[1]{\hat{\lambda}_{#1}}
\newcommand{\logLamb}{\hat{\mat{\Lambda}}}
\newcommand{\logEffShear}{\hat{G}_\mathrm{eff}}
\newcommand{\effShear}{G_\mathrm{eff}}
\newcommand{\bludShear}{G_\mathrm{B}}

\newcommand{\x}{\vec{x}}
\newcommand{\surfStrain}{\varepsilon}
\newcommand{\charTime}{\tau}
\newcommand{\poreArea}{A_\mathrm{p}}
\newcommand{\poreRadius}{r_\mathrm{p}}

\newcommand{\effStress}{\sigma_\mathrm{eff}}
\newcommand{\qof}{\vec{\phi}}

\definecolor{darkblue}{RGB}{  0  84 160}
\definecolor{petrol}{RGB}{  0  98 101}
\definecolor{bordeaux}{RGB}{161  17  53}

\newacronym{CFD}{CFD}{computational fluid dynamics}
\newacronym[prefixfirst={a\ },
            prefix={an\ }]{RBC}{RBC}{red blood cell}
\newacronym[prefixfirst={a\ },
            prefix={an\ }]{LES}{LES}{large eddy simulation}
\newacronym{VMS}{VMS}{variational multiscale}
\newacronym{RANS}{RANS}{Reynolds-averaged Navier-Stokes}
\newacronym{IH}{IH}{index of hemolysis}
\newacronym{TTM}{TTM}{tank-treading model}
\newacronym{TTLM}{TTLM}{logarithmic tank-treading model}
\newacronym{KV}{KV}{Kelvin-Voigt}
\newacronym{KVL}{KVL}{logarithmic Kelvin-Voigt}
\newacronym[prefixfirst={an\ },
            prefix={a\ }]{ODE}{ODE}{ordinary differential equation}
\newacronym[prefixfirst={a\ },
            prefix={an\ }]{MRF}{MRF}{multiple reference frame}
\newacronym{PDE}{PDE}{partial differential equation}
\newacronym{FDA}{FDA}{U.S. Food and Drug Administration}
\newacronym{PIV}{PIV}{particle image velocimetry}
\newacronym{MIH}{MIH}{modified index of hemolysis}

\begin{document}

\def\titlename{A Practical Computational Hemolysis Model Incorporating Biophysical Properties of the Red Blood Cell Membrane}

\title[\titlename]{\titlename}

\author*[1]{\fnm{Nico} \sur{Dirkes}}\email{dirkes@aices.rwth-aachen.de}

\author[1]{\fnm{Marek} \sur{Behr}}

\affil[1]{\orgdiv{Chair for Computational Analysis of Technical Systems}, \orgname{RWTH Aachen University}, \orgaddress{\street{Schinkelstr. 2}, \city{Aachen}, \postcode{52062}, \country{Germany}}}

\abstract{\textbf{Purpose:} Hemolysis is a key issue in the design of blood-handling medical devices. Computational prediction of this phenomenon is challenging due to the complex multiscale nature of blood. As a result, conventional approaches often fail to predict hemolysis accurately, commonly showing deviations of multiple orders of magnitude compared to experimental data. More accurate models are typically computationally expensive and thus impractical for real-world applications. This work aims to fill this gap by presenting accurate yet simple and efficient computational hemolysis models.

\textbf{Methods:} Hemolysis modeling relies on two key components: a red blood cell model and a hemoglobin release model. In this work, we compare three red blood cell models: a common stress-based model (Bludszuweit), a simple strain-based model based on the Kelvin-Voigt constitutive law, and a more complex tensor-based model (TTM). Further, we compare two hemoglobin release models: the widely used power-law approach and a biophysical pore formation model. 

\textbf{Results:} We evaluate these models in two benchmark cases: the FDA blood pump and the FDA nozzle. In both benchmarks, the simple strain-based model combined with the pore formation model achieves absolute predictions of hemolysis within the standard deviation of experimental measurements. In contrast, stress-based power law models deviate by several orders of magnitude. 

\textbf{Conclusion:} The strain-based pore modeling approach takes into account the biophysical properties of red blood cell membranes, in particular their viscoelastic deformation behavior and hemoglobin release through membrane pores. This leads to significantly improved hemolysis predictions in a framework that can easily be integrated into common CFD workflows.
}


\maketitle

\section{Introduction}
\label{sec:introduction}

Computational analysis has become an indispensable tool in the design process of blood-handling medical devices, such as ventricular assist devices, as it allows to predict device performance prior to manufacturing and testing. Two particular areas of interest are hydraulic performance, quantifying the pressure head and flow rate, and hematologic performance, quantifying the damage to blood components, such as \glspl{RBC}. The hematologic performance consists of multiple aspects, among which \emph{hemolysis}, i.e., flow-induced \gls{RBC} damage, is of particular clinical relevance~\cite{katzMulticenterAnalysisClinical2015c}. In hemolysis, fluid stresses distort 
the \gls{RBC}, which can lead to membrane poration and ultimately rupture. As a result, the hemoglobin contained within the \gls{RBC} leaks out into blood plasma, which can lead to severe complications, such as renal failure, bleeding and thrombosis~\cite{shahBleedingThrombosisAssociated2017,omarPlasmaFreeHemoglobin2015,lyuPlasmaFreeHemoglobin2016}.
While \gls{CFD} has matured to a reliable tool for predicting hydraulic performance, the accurate prediction of hemolysis remains challenging. In a recent interlaboratory study~\cite{ponnaluriResultsInterlaboratoryComputational2023b}, hydraulic predictions were generally within a range of $\pm 20\%$ of experimental values, whereas hemolysis predictions varied by multiple orders of magnitude. This highlights the lack of reliable hemolysis models and calls for the development of improved approaches.

The reason hemolysis prediction is challenging lies in the complex multiscale nature of the problem. Hemolysis occurs on the scale of individual \glspl{RBC}, which have a diameter of approximately $8\, \unit{\micro\meter}$. In contrast, blood-handling medical devices have characteristic dimensions on the order of centimeters, i.e., four orders of magnitude larger. Plenty of detailed numerical methods exist to resolve the shape of individual \glspl{RBC}~\cite{ezzeldinStrainBasedModelMechanical2015b,zavodszkyCellularLevelInSilico2017b,mendezUnstructuredSolverSimulations2014,fedosovMultiscaleRedBlood2010b,gugliettaLoadingRelaxationDynamics2021b}. These methods can accurately capture the biophysical properties of the \gls{RBC} membrane and support insights into the underlying mechanisms of hemolysis. However, their computational cost limits the number of \glspl{RBC} that can be simulated. As a result, these methods typically track only a small fraction of all \glspl{RBC} flowing through a device, evaluating hemolysis along their individual trajectories. 
Such Lagrangian approaches have two main drawbacks. First, it is unclear how to select a representative sample of \glspl{RBC} to achieve accurate predictions. Second, the selected \glspl{RBC} may not penetrate all relevant regions of the flow domain, such as small gap flows, boundary layers, and recirculation zones. Since these regions are most critical for hemolysis, resolving them accurately is essential for a model's utility in device design. For these reasons, such Lagrangian hemolysis models are not well suited to support the design process of blood-handling medical devices.

Eulerian models, on the other hand, treat blood as a continuum and describe the average behavior of \glspl{RBC} at each point in space and time.  This gives a spatially resolved field for \gls{RBC} deformation, allowing for easy identification of critical regions inside the domain. However, developing accurate Eulerian hemolysis models is challenging.
The models need to incorporate microscale behavior of \glspl{RBC} into effective constitutive equations on the macroscale. 
In this work, we will focus on modeling two important biophysical properties of the \gls{RBC} membrane: the viscoelastic deformation in response to fluid stress and the formation of pores as physical mechanism for hemoglobin release. 

The viscoelastic nature of the \gls{RBC} membrane has been documented extensively in literature~\cite{puig-de-morales-marinkovicViscoelasticityHumanRed2007b,hochmuthRedCellExtensional1979,katchalskyRheologicalConsiderationsHaemolysing1960,randMechanicalPropertiesRed1964}. 
Conventional hemolysis models, so-called \emph{stress-based} models, often neglect this property and assume that \glspl{RBC} deform instantaneously to their steady state when encountering fluid stress. However,
the importance of incorporating viscoelastic effects into hemolysis models has recently been highlighted again experimentally~\cite{lommelExperimentalInvestigationApplicability}. So-called \emph{strain-based} hemolysis models incorporate viscoelastic \gls{RBC} deformation models to account for the characteristic response of \glspl{RBC} to fluid stress.
The simplest form of a viscoelastic model is the scalar \gls{KV} model, which consists of a spring and a dashpot in parallel. It does not explicitly resolve the membrane structure but provides a scalar measure of its deformation over time. It captures both the elastic and viscous response of the membrane to fluid stress. This model has first been suggested in the context of \gls{RBC} mechanics by \citet{katchalskyRheologicalConsiderationsHaemolysing1960}. Afterwards, \citet{randMechanicalPropertiesRed1964} extended the model by an additional spring and dashpot in series to capture the flowing behavior of \glspl{RBC}. More recently, this approach has been extended to directly model the plasma-free hemoglobin~\cite{arwatzViscoelasticModelShearInduced2013c,chenStrainBasedFlowInducedHemolysis2011a,yeleswarapuMathematicalModelShearInduced1995}. However, these extended forms of the \gls{KV} model are generally only valid for specific flow conditions, such as constant shear rates~\cite{arwatzViscoelasticModelShearInduced2013c} or Poiseuille flow~\cite{chenStrainBasedFlowInducedHemolysis2011a}. Extensions to more general three-dimensional flow fields have been attempted with mixed success~\cite{chenTestingModelsFlowInduced2013c,leeEvaluationExtendedViscoelastic2019,yuReviewHemolysisPrediction2017c}. In addition, these models are scalar in nature and do not consider the effect of three-dimensional stress states on \gls{RBC} deformation. For this reason, we will focus on two particular models: the \gls{KV} model and \gls{TTM}~\cite{dirkesEulerianFormulationTensorbased2024}. While the \gls{KV} model resolves viscoelastic deformation in a scalar manner, the \gls{TTM} approximates \glspl{RBC} as three-dimensional ellipsoids and captures their orientation in space. Thus, it accounts for the effect of three-dimensional stress states on \gls{RBC} deformation. In particular, this can account for the difference between shear and extensional stress~\cite{dirkesSignificanceFlowVorticity2025,faghihDeformationHumanRed2020b}. The comparison of these two models will allow us to assess the importance of resolving three-dimensional stress states for hemolysis prediction.

The formation of pores in the \gls{RBC} membrane under excessive membrane strain causes hemoglobin release through diffusion and advection. This mechanism is assumed to be the primary mechanism for sublethal hemolysis~\cite{faghihModelingPredictionFlowinduced2019a}. In contrast, lethal hemolysis occurs when the membrane ruptures entirely, releasing all hemoglobin contained within the \gls{RBC} at once. While there have been attempts to combine models for sublethal and lethal hemolysis~\cite{poorkhalilNewApproachSemiempirical2016,mckeanDevelopmentHemolysisModel2020}, the additional complexity was overall deemed to be of limited benefit. Thus, we will focus on modeling sublethal hemolysis only. 
Historically, hemoglobin release has often been modeled using empirical power-law correlations that relate fluid stress and exposure time to hemolysis~\cite{giersiepenEstimationShearStressRelated1990a,dingShearInducedHemolysisSpecies2015a}. However, such models are purely empirical and do not incorporate the underlying biophysical mechanisms of hemolysis. In contrast, pore formation models describe hemoglobin release based on the number and size of pores formed in the \gls{RBC} membrane as a function of membrane strain. \citet{vitaleMultiscaleBiophysicalModel2014a} presented a first biophysical model for pore formation and hemoglobin release, assuming a uniform pore distribution on the membrane. The model is applicable in both Eulerian and Lagrangian frameworks. Afterwards, \citet{sohrabiCellularModelShearInduced2017b} developed a more complex model with non-uniform pore distribution for more general shape distortions. Due to the increased complexity, the model is only applicable in Lagrangian formulation. We will employ the simpler model by \citet{vitaleMultiscaleBiophysicalModel2014a} in the present work, as it is more suitable for our Eulerian approach, and has been shown to achieve satisfactory results in previous studies~\cite{hasslerFiniteelementFormulationAdvection2020a}.

The novelty in our approach thus lies in the combination of the aforementioned aspects: We aim to find a simple macroscopic Eulerian hemolysis model that captures viscoelastic effects and pore formation. This combination results in a computationally efficient hemolysis model that incorporates important biophysical properties of the \gls{RBC} membrane and provides spatially resolved fields to aid the design process of medical devices. It should be easy to implement in existing \gls{CFD} solvers, and be applicable to general three-dimensional flow fields for large-scale simulations of blood-handling medical devices. We further aim to validate our model against experimental data for relevant benchmark problems to prove its accuracy and reliability. For comparison, we evaluate different combinations of stress-based and strain-based approaches with empirical power law and pore-based hemoglobin release models, providing a comprehensive assessment of the different modeling choices.

This paper is structured as follows: In Section~\ref{sec:methods}, we present the three fundamental components of our computational hemolysis framework, namely \gls{CFD}, \gls{RBC} model, and hemoglobin release model. We present different choices for each component and discuss their advantages and disadvantages. We show how we evaluate these models numerically. In Section~\ref{sec:results}, we present results for two benchmark problems defined by the \gls{FDA}: a rotary blood pump and a nozzle. We evaluate different combinations of \gls{RBC} models and hemoglobin release models and compare them to experimental data. In Section~\ref{sec:discussion}, we discuss the results and their implications for future research.

\section{Methods}
\label{sec:methods}

\begin{figure}
    \centering
    \includegraphics[width=\textwidth]{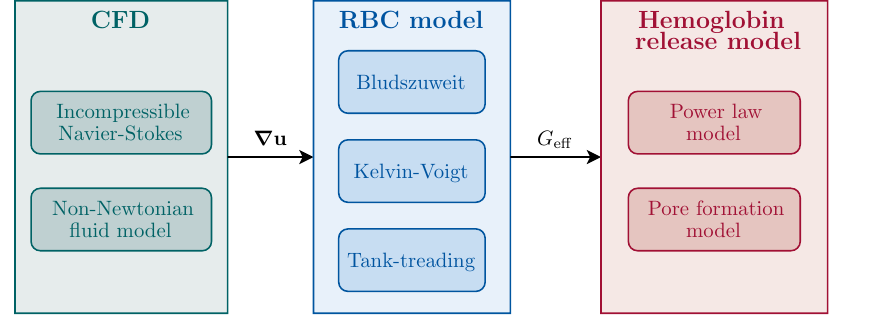}
    \caption{Hemolysis model components.}
    \label{fig:hemolysis_modeling_overview}
\end{figure}

Hemolysis models generally consist of three components: \gls{CFD} to determine the flow field, \pgls{RBC} model to evaluate cell deformation, and a hemoglobin release model to quantify plasma-free hemoglobin. The overall workflow is illustrated in \cref{fig:hemolysis_modeling_overview}. This section outlines the different modeling choices for these components and the numerical methods we use to solve them. 

\subsection{Computational Fluid Dynamics}
\label{sec:methods_cfd}

In this work, we model blood as an incompressible Newtonian fluid. The governing equations are the continuity equation and the Navier-Stokes equations:
\begin{equation}
\begin{aligned}
    \grad \cdot \vel &= 0 \ , \\
    \rho \left[ \matDer{\vel} \right] &= \mu \grad \cdot \strain - \grad p \ , \qquad
    \strain = \symVelGrad \ ,
\end{aligned}
\label{eq:navier_stokes}
\end{equation}
Here, $\vel$ is the fluid velocity, $p$ is the pressure, $\rho$ is the density, $\mu$ is the dynamic viscosity, and $\strain$ is the rate of strain tensor. The Newtonian fluid assumption is commonly used in large vessels and at high shear rates \cite{pauliStabilizedFiniteElement2016c}. More sophisticated fluid models exist to account for the non-Newtonian behavior of blood \cite{melkaNumericalInvestigationMultiphase2019,zavodszkyCellularLevelInSilico2017b,bodnarSimulationThreeDimensionalFlow2011}. In this work, we will focus on the Newtonian assumption. However, all presented \gls{RBC} and hemoglobin models may easily be combined with non-Newtonian fluid models. Independent of the employed fluid model, the \gls{CFD} generally only needs to provide the velocity field $\vel(\x, t)$ and the velocity gradient $\grad \vel(\x, t)$ to the \gls{RBC} model. 

For viscoelastic models, the fluid stresses are typically split into a solvent and polymeric contribution. In that case, the solvent contribution is based on the rate of strain tensor $\strain$, while the polymeric contribution is based on additional equations that model the underlying microstructure of the fluid on the macroscopic scale. The \gls{RBC} models presented in the following section incorporate similar microstructural effects in a one-way coupling. The most direct approach to couple them to viscoelastic fluid models is thus to simply use the full velocity gradient $\grad \vel$ from a viscoelastic \gls{CFD} simulation in the same one-way coupling. A more sophisticated approach is to directly integrate the \gls{RBC} model into the viscoelastic fluid model, resulting in a two-way coupling. This is the subject of ongoing research.

\subsection{Red Blood Cell Models}
\label{sec:methods_rbc_models}

\begin{figure}
    \centering
    \begin{subfigure}[t]{0.48\textwidth}
        \vskip 2em
        \includegraphics[width=0.8\textwidth]{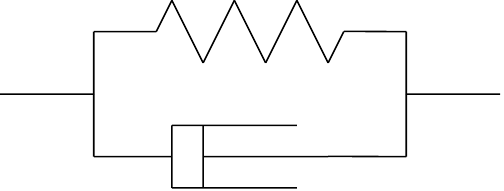}
        \vskip 3.25em
        \caption{Schematic of the \gls{KV} model consisting of a spring and dashpot in parallel.}
        \label{fig:kelvin_voigt_schematic}
    \end{subfigure}
    \hfill
    \begin{subfigure}[t]{0.48\textwidth}
        \vskip 0pt
        \includegraphics{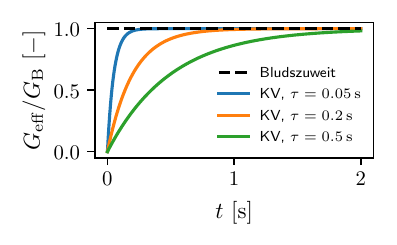}
        \vskip -1em
        \caption{Response of stress-based Bludszuweit model and strain-based \gls{KV} model to a step increase in fluid shear rate $\bludShear$ for different relaxation times $\charTime$.}
        \label{fig:kelvin_voigt_response}
    \end{subfigure}
    \begin{subfigure}{0.48\textwidth}
        \includegraphics{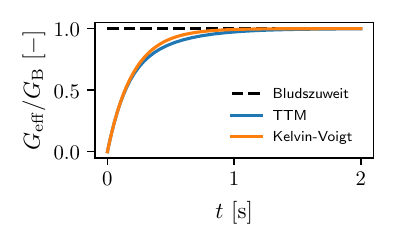}
        \caption{Response of \gls{RBC} models to step increase in simple shear flow.}
        \label{fig:ttm_response_shear}
    \end{subfigure}
    \hfill
    \begin{subfigure}{0.48\textwidth}
        \includegraphics{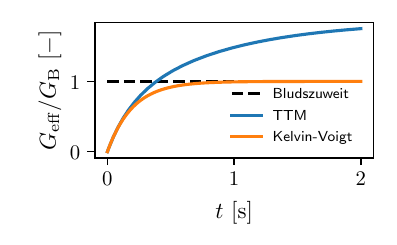}
        \caption{Response of \gls{RBC} models to step increase in extensional flow.}
        \label{fig:ttm_response_extensional}
    \end{subfigure}
    \begin{subfigure}{\textwidth}
        \includegraphics[width=\textwidth]{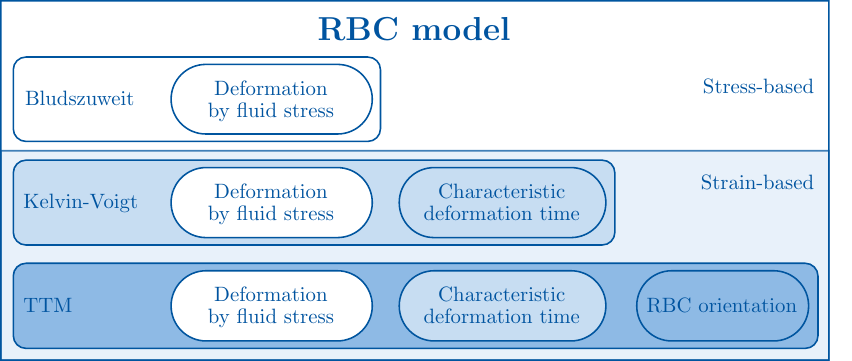}
        \caption{Overview of effects taken into account by the three different \gls{RBC} models.}
        \label{fig:rbc_models}
    \end{subfigure}
    \caption{Illustration of the differences between the three types of \gls{RBC} models employed in this work.}
    \label{fig:rbc_models_all}
\end{figure}

We consider three different \gls{RBC} models to evaluate cell deformation in response to the flow field obtained from the \gls{CFD} simulation: the Bludszuweit model, the \acrfull{TTM}, and the \acrfull{KV} model. These models differ in complexity and the biophysical properties they account for. An overview of the differences between the models is given in \cref{fig:rbc_models_all}. All models employ the concept of an \emph{effective shear rate} $\effShear$ to quantify cell deformation. This quantity represents the equivalent fluid shear rate that would lead to the same deformation at steady state under simple shear.
The effective shear rate is then used in the hemoglobin release model to quantify hemolysis. We employ the effective shear rate instead of the effective shear stress $\effStress = \mu \effShear$ in order to achieve independence from the fluid viscosity $\mu$. In fact, a recent study~\cite{krisherEffectBloodViscosity2022} has found that hemolysis depends primarily on the shear rate rather than the shear stress.

First, we define the Bludszuweit shear rate $\bludShear$ in analogy to the representative stress proposed by Bludszuweit~\cite{bludszuweitModelGeneralMechanical1995c}:
\begin{equation}
    \bludShear = 
    \frac{2}{\sqrt{3}} \sqrt{ 
        \strainI{xx}^2 + \strainI{yy}^2 + \strainI{zz}^2
        - \strainI{xx}\strainI{yy} - \strainI{yy}\strainI{zz} - \strainI{zz}\strainI{xx}
        + 3(\strainI{xy}^2 + \strainI{yz}^2 + \strainI{zx}^2)
    } \, .
    \label{eq:blud_shear}
\end{equation}
This serves as a scalar measure of the three-dimensional rate of strain tensor $\strain$. The \emph{stress-based} approach consists of setting the effective shear rate $\effShear$ equal to the Bludszuweit shear rate:
\begin{equation}
    \effShear = \bludShear \, .
    \label{eq:effShear_blud}
\end{equation}
As \cref{fig:kelvin_voigt_response} illustrates, this corresponds to an instantaneous elastic response of the \gls{RBC} to fluid stress. 

Next, we consider the \emph{strain-based} modeling approach. Models in this category take into account the viscoelastic properties of the \gls{RBC} membrane~\cite{puig-de-morales-marinkovicViscoelasticityHumanRed2007b}. The viscoelastic response has been shown to be essential in modeling the effect of short-term stresses \cite{lommelExperimentalInvestigationApplicability}.
The simplest way to account for this is using a zero-dimensional \gls{KV} model for a viscoelastic solid, namely a spring and dashpot in parallel. The spring governs the elastic response and limits the deformation under constant load. The dashpot governs the viscous response and introduces a time delay in the deformation. This scheme is illustrated in \cref{fig:kelvin_voigt_schematic}. If the deformation of the \gls{RBC} is expressed by the effective shear rate $\effShear$, the \gls{KV} model results in the following transport equation:
\begin{equation}
    \mDDt{\effShear} = F^\mathrm{KV}(\effShear; \grad\vel) \coloneqq \frac{1}{\charTime} \left( \bludShear - \effShear \right) \, ,
    \label{eq:effShear_kv}
\end{equation}
where $\charTime$ is the relaxation time of the material, which governs the timescale of the viscoelastic response to changes in fluid stress. The effect of this parameter is illustrated in \cref{fig:kelvin_voigt_response}. The smaller the timescale $\charTime$, the faster the response to a sudden change in fluid stress. For $\charTime \to 0$, the \gls{KV} model tends to the Bludszuweit model~\eqref{eq:effShear_blud}, as the response becomes instantaneous. In that sense, it extends the purely elastic stress-based model by viscoelastic effects. In this work, we set $\charTime = \SI{200}{\milli\second}$ based on experimental data~\cite{henonNewDeterminationShear1999,bronkhorstNewMethodStudy1995}.

Finally, we present the \acrfull{TTM}~\cite{dirkesEulerianFormulationTensorbased2024}. It approximates red blood cells as ellipsoids. Each ellipsoid is described in terms of its semi-axes:
\begin{equation}
    \Lamb = \diag (\lambda_1, \lambda_2, \lambda_3) \, , \qquad
    \Qm = \begin{bmatrix}
        \vec{v}_1 & \vec{v}_2 & \vec{v}_3
    \end{bmatrix} \, .
\end{equation}
The length of the $i$-th semi-axis is given by $\sqrt{\lambda_i}$, and the normalized direction of the $i$-th semi-axis is given by the vector $\vec{v}_i$. The matrix $\Qm$ thus describes the orientation of the ellipsoid in space. 
The model is written as a transport equation for $\Lamb$:
\begin{subequations}\label{eq:ttm}
\begin{align}
    \mDDt{\Lamb} = \mathbf{F}^\mathrm{TTM}(\Lamb; \grad\vel) 
    \, , \quad 
    F^\mathrm{TTM}_i(\Lamb; \grad\vel)  = -f_1 \Big[ \lambda_i - g(\Lamb) \Big] + 2 f_2 \lambda_i \straintI{ii} \, , 
    \label{eq:ttm_deformation} \\
    \straint = \trafoCoord{\strain} \, , \quad
    \Qm = 
    \begin{dcases}
        \Qm_\star(\grad\vel, \Lamb) \, , \quad &\text{tank-treading}\, , \\
        \mat{0} \, , \quad &\text{tumbling}\, .
    \end{dcases} \label{eq:ttm_orientation}
\end{align} 
We choose $f_1 = \frac{1}{\charTime} = \SI{5}{\per\second}$ and $f_2 = 4.2298\cdot 10^{-4}$ to match relaxation time and surface strain of experimental data, see~\cite{aroraTensorBasedMeasureEstimating2004}. \Cref{eq:ttm_deformation} models the viscoelastic deformation of the \gls{RBC} with the same relaxation time $\charTime$ as the \gls{KV} model. \Cref{eq:ttm_orientation} models the orientation dynamics of the \gls{RBC}, accounting for tank-treading and tumbling motions depending on the flow conditions $\grad\vel$ and deformation $\Lamb$. The orientation equation can be solved in an iterative algorithm. 
Assuming that $\lambda_1 > \lambda_2 > \lambda_3$, the effective shear rate in this model is computed from the deformation of the ellipsoid as
\begin{equation}
    \effShear = \frac{2 D f_1}{(1 - D^2) f_2} \, , \qquad
    D = \frac{\sqrt{\lambda_1} - \sqrt{\lambda_3}}{\sqrt{\lambda_1} + \sqrt{\lambda_3}} \, .
    \label{eq:ttm_effShear}
\end{equation}
\end{subequations}
The differences between the three models are illustrated in \cref{fig:ttm_response_shear,fig:ttm_response_extensional}. In simple shear flow, the \gls{TTM} and \gls{KV} model are practically identical, as both models predict viscoelastic response with a characteristic relaxation time of \SI{200}{\milli\second}. In extensional flow, however, the \gls{TTM} predicts a significantly higher deformation. This is due to the alignment of the red blood cell axes with the principal axes of fluid stress, and has been observed experimentally~\cite{faghihDeformationHumanRed2020b}. For more details on the incorporation of this effect in the \gls{TTM}, see also \cite{dirkesSignificanceFlowVorticity2025}.
An overview of the three presented \gls{RBC} models is given in \cref{fig:rbc_models}.

\bigskip
Having introduced the defining features of the three \gls{RBC} models employed in this work, we now define quantities to compare them systematically.
First, we note that the \gls{TTM} requires a logarithmic formulation for stability in complex flows, in analogy to other tensor-based models~\cite{dirkesEulerianFormulationTensorbased2024,hasslerVariationalMultiscaleFormulation2019b}:
\begin{equation}
    \mDDt{\logLamb} = \mathbf{F}^\mathrm{TTLM}(\logLamb; \grad\vel) \, .
    \label{eq:ttlm}
\end{equation}
We call this the \gls{TTLM}. We analogously write the \gls{KV} model in logarithmic form for consistency:
\begin{equation}
    \mDDt{\logEffShear} = F^\mathrm{KVL}(\logEffShear; \grad\vel) \coloneqq 
    \bludShear / \effShear - 1 \, , \qquad
    \effShear = e^{\logEffShear / \charTime} \, .
    \label{eq:effShear_kvl}
\end{equation}
We call this the \gls{KVL} model. 
In order to evaluate the effect of the logarithmic formulation, we define the relative difference in effective shear rates predicted by the \gls{KVL} and the original \gls{KV} model:
\begin{equation}
    \delta_{G}^{\mathrm{KV};\mathrm{KVL}} = \frac{\effShear^\mathrm{KV} - \effShear^\mathrm{KVL}}{\effShear^\mathrm{KVL}} \, .
    \label{eq:deltaG_kv_kvl}
\end{equation}
The \gls{KVL} further serves as a basis for comparison with the \gls{TTLM}, as both models employ a logarithmic formulation.
We compute the relative difference in effective shear rates predicted by the two models:
\begin{equation}
    \delta_{G}^{\mathrm{KVL};\mathrm{TTLM}} = \frac{\effShear^\mathrm{KVL} - \effShear^\mathrm{TTLM}}{\effShear^\mathrm{TTLM}} \, .
    \label{eq:deltaG_kvl_ttlm}
\end{equation}
To further analyze the effect of the three-dimensional fluid stress and orientation, we define a shear-only version of the \gls{TTLM}. This version reduces the three-dimensional velocity gradient to a simple shear flow, using the Bludszuweit shear rate~\eqref{eq:blud_shear}:
\begin{equation}
    \mDDt{\Lamb^\mathrm{s}} = \mathbf{F}^\mathrm{TTLM}(\Lamb^\mathrm{s} ; \grad\vel^\mathrm{s}) \, , \qquad
    \grad\vel^\mathrm{s} \coloneqq \begin{pmatrix}
        0 & \bludShear (\grad\vel) & 0 \\
        0 & 0 & 0 \\
        0 & 0 & 0
    \end{pmatrix} \, .
    \label{eq:ttlm_shear}
\end{equation}
Here, $\grad\vel^\mathrm{s}$ imitates the velocity gradient in simple shear flow of intensity $\bludShear$. We then compute the effective shear rate $\effShear^\mathrm{s}$ from the deformation $\Lamb^\mathrm{s}$ of this shear-only model as in \cref{eq:ttm_effShear}. Overall, this version of the \gls{TTLM} is equivalent to the \gls{KVL} model, in that it considers only viscoelastic deformation through the magnitude of $\bludShear$, without any three-dimensional effects. The main difference is that this formulation allows us to compare more directly to the full \gls{TTLM}, since it has the same equation structure in $\Lamb$. We thus employ this model to compute the relative differences in the largest eigenvalue and in the respective source term $F_1$:
\begin{equation}
    \delta_\Lambda = \frac{{\eigval_1} - {\eigval_1^\mathrm{s}}}{{\eigval_1^\mathrm{s}}} \, ,
    \qquad
    \delta_F = \frac{F_1^\mathrm{TTLM}(\Lamb ; \grad\vel) - F_1^\mathrm{TTLM}(\Lamb^\mathrm{s} ; \grad\vel^\mathrm{s})}{F_1^\mathrm{TTLM}(\Lamb^\mathrm{s} ; \grad\vel^\mathrm{s})} \, .
    \label{eq:deltaL_deltaF}
\end{equation}
On the one hand, $\delta_\Lambda$ quantifies the influence of the three-dimensional \gls{RBC} orientation on the predicted \emph{deformation}. On the other hand, $\delta_F$ quantifies the influence of 3D \gls{RBC} orientation on the predicted \emph{rate of deformation}.

\subsection{Hemoglobin Release Models}
\label{sec:methods_hemoglobin_models}
Next, we present two choices for models to quantify hemoglobin release from damaged \glspl{RBC}. Both models rely on the effective shear rate $\effShear$ computed from the \gls{RBC} models in \cref{sec:methods_rbc_models}, and compute the local index of hemolysis $IH$ as a measure of increase in plasma-free hemoglobin concentration $\mathrm{Hb}$:
\begin{equation}
    IH(\x, t) = \frac{\mathrm{Hb}(\x, t) - \mathrm{Hb}_0}{\mathrm{Hb}_0} \, .
\end{equation}
To compare with experimental data, we report the modified index of hemolysis $MIH$ as a global quantity by integrating $IH$ over the outlet of the device $\Gamma_\text{out}$:
\begin{equation}
    MIH = \frac{10^6}{Q} \int_{\Gamma_\text{out}} IH(\x, t) \vel(\x, t) \cdot \vec{n} \, \mathrm{d}A \, , \qquad
    Q = 
    \int_{\Gamma_\text{out}} \vel(\x, t) \cdot \vec{n} \, \mathrm{d}A \, .
    \label{eq:mih}
\end{equation}

First, we present the widely used power law model, originally proposed by \citet{giersiepenEstimationShearStressRelated1990a}:
\begin{equation}
    IH = A \effStress^\alpha t^\beta \, ,
    \qquad 
    \effStress = \mu \effShear \, .
    \label{eq:power_law_algebraic}
\end{equation}
It relates the index of hemolysis $IH$ to the effective fluid stress $\effStress$ and the exposure time $t$. The parameters $A$, $\alpha$, and $\beta$ are fitted to experimental data. In the form of~\cref{eq:power_law_algebraic}, the model is only valid for constant effective shear rates. To model varying shear rates, there are various strategies, see~\cite{taskinEvaluationEulerianLagrangian2012b}. 
We use the raw hemolysis data from \citet{dingShearInducedHemolysisSpecies2015a} to fit the parameters for four different species: human, ovine, porcine, and bovine. For the stress-based model~\eqref{eq:effShear_blud}, we directly use the parameters reported by \citet{dingShearInducedHemolysisSpecies2015a}. For the strain-based models~\labelcref{eq:effShear_kv,eq:ttm}, we use the analytical solution of the \gls{KV} model for constant fluid shear rates to compute the effective shear rate over time. We then employ the linearized form of the power law model to compute the index of hemolysis:
\begin{equation}
    \ell_{IH} = IH^\frac{1}{\beta} \, , \quad
    \mDDt{\ell_{IH}} = A^\frac{1}{\beta} \effStress^\frac{\alpha}{\beta}(t) \, , 
    \quad
    \effStress(t) = \mu \effShear(t) \, , \quad
    \effShear(t) = \bludShear \left( 1 - e^{-t/\charTime} \right) \, .
    \label{eq:effShear_kv_analytical}
\end{equation}
The parameters $(A, \alpha, \beta)$ for the strain-based power law are determined by fitting the results to the raw experimental data of \citet{dingShearInducedHemolysisSpecies2015a}. For this purpose, the \gls{ODE} for $\ell_{IH}$ is integrated numerically with Python using \texttt{scipy.integrate.quad}. The parameter fitting is performed using \texttt{scipy.optimize.curve\_fit}. 
The resulting parameters are summarized in \cref{tab:power_law_parameters}. In analogy to \citet{dingShearInducedHemolysisSpecies2015a}, we report the correlation coefficient $R$ as a measure of goodness of fit.
\JSONParseFromFile{\dingPowerLawParams}{params_powerStress.json}
\JSONParseValue[rescan, store in=\powerLawStressOvineA]{\dingPowerLawParams}{ovine.A}
\JSONParseValue[rescan, store in=\powerLawStressBovineA]{\dingPowerLawParams}{bovine.A}
\JSONParseValue[rescan, store in=\powerLawStressPorcineA]{\dingPowerLawParams}{porcine.A}
\JSONParseValue[rescan, store in=\powerLawStressHumanA]{\dingPowerLawParams}{human.A}
\JSONParseFromFile{\strainPowerLawParams}{params_powerStrain.json}
\JSONParseValue[rescan, store in=\powerLawStrainOvineA]{\strainPowerLawParams}{ovine.A}
\JSONParseValue[rescan, store in=\powerLawStrainBovineA]{\strainPowerLawParams}{bovine.A}
\JSONParseValue[rescan, store in=\powerLawStrainPorcineA]{\strainPowerLawParams}{porcine.A}
\JSONParseValue[rescan, store in=\powerLawStrainHumanA]{\strainPowerLawParams}{human.A}
\begin{table}
    \centering
    \begin{tabular}{ll
        S[table-format=1.3e-1]
        S[table-format=1.4]
        S[table-format=1.4]
        S[table-format=1.4]}
        \toprule
        \gls{RBC} Model & Species & {$A$} & {$\alpha$} & {$\beta$} & {Correlation coefficient ($R$)} \\
        \midrule
        stress-based & ovine & 
        \powerLawStressOvineA
        & \JSONParseExpandableValue{\dingPowerLawParams}{ovine.alpha} 
        & \JSONParseExpandableValue{\dingPowerLawParams}{ovine.beta} 
        & \JSONParseExpandableValue{\dingPowerLawParams}{ovine.r} 
        \\
        & human & 
        \powerLawStressHumanA & 
        \JSONParseExpandableValue{\dingPowerLawParams}{human.alpha} & 
        \JSONParseExpandableValue{\dingPowerLawParams}{human.beta} &
        \JSONParseExpandableValue{\dingPowerLawParams}{human.r}
        \\
        & porcine & 
        \powerLawStressPorcineA & 
        \JSONParseExpandableValue{\dingPowerLawParams}{porcine.alpha} & 
        \JSONParseExpandableValue{\dingPowerLawParams}{porcine.beta} &
        \JSONParseExpandableValue{\dingPowerLawParams}{porcine.r}
        \\
        & bovine & 
        \powerLawStressBovineA & 
        \JSONParseExpandableValue{\dingPowerLawParams}{bovine.alpha} & 
        \JSONParseExpandableValue{\dingPowerLawParams}{bovine.beta} &
        \JSONParseExpandableValue{\dingPowerLawParams}{bovine.r}
        \\[1ex]
        strain-based & ovine & 
        \powerLawStrainOvineA &
        \JSONParseExpandableValue{\strainPowerLawParams}{ovine.alpha} & 
        \JSONParseExpandableValue{\strainPowerLawParams}{ovine.beta} &
        \JSONParseExpandableValue{\strainPowerLawParams}{ovine.r}
        \\
        & human & 
        \powerLawStrainHumanA & 
        \JSONParseExpandableValue{\strainPowerLawParams}{human.alpha} & 
        \JSONParseExpandableValue{\strainPowerLawParams}{human.beta} &
        \JSONParseExpandableValue{\strainPowerLawParams}{human.r}
        \\
        & porcine & 
        \powerLawStrainPorcineA & 
        \JSONParseExpandableValue{\strainPowerLawParams}{porcine.alpha} & 
        \JSONParseExpandableValue{\strainPowerLawParams}{porcine.beta} &
        \JSONParseExpandableValue{\strainPowerLawParams}{porcine.r}
        \\
        & bovine & 
        \powerLawStrainBovineA & 
        \JSONParseExpandableValue{\strainPowerLawParams}{bovine.alpha} & 
        \JSONParseExpandableValue{\strainPowerLawParams}{bovine.beta} &
        \JSONParseExpandableValue{\strainPowerLawParams}{bovine.r}
        \\
        \bottomrule
    \end{tabular}
    \caption{Power law parameters for different species and \gls{RBC} models, fitted to experimental data of \citet{dingShearInducedHemolysisSpecies2015a}.}
    \label{tab:power_law_parameters}
\end{table}

There are a number of drawbacks to the power law model. First, the model is purely empirical and does not account for the underlying biophysical mechanisms of hemolysis. Second, it has been shown that many choices of parameters $(A, \alpha, \beta)$ can fit the same experimental data almost equally well, leading to large variations in the determined hemolysis parameters~\cite{blumQuantifyingExperimentalVariability2025}. Third, the extension to non-constant fluid stress histories is not always valid. In particular for rapidly changing shear rates, the linearization can induce significant errors~\cite{faghihPracticalImplicationsErroneous2023}. However, the approach is frequently used in practice due to its simplicity~\cite{lommelExperimentalInvestigationApplicability,lacasseMechanicalHemolysisBlood2007,cravenCFDbasedKrigingSurrogate2019,yuReviewHemolysisPrediction2017c}. Extensions have been proposed to include threshold and saturation effects~\cite{yuReviewHemolysisPrediction2017c}. 

Second, we present the pore formation model proposed by \citet{vitaleMultiscaleBiophysicalModel2014a}. Compared to the original model, we employ a simplified form:
\begin{subequations}\label{eq:pore_formation}
\begin{equation}
    \mDDt{IH} = h \sigma_\mathrm{B}^k \poreArea(\effShear) \, ,
    \qquad 
    \sigma_\mathrm{B} = \mu G_\mathrm{B} \, .
    \label{eq:pore_formation_mddt}
\end{equation}
The model relates hemoglobin release to the formation of pores in the membrane. Compared to the original model, the factor $1/V_\mathrm{RBC}$ has been absorbed into the parameter $h$, and the saturation term is dropped assuming $IH \ll 1$. Furthermore, we use the Bludszuweit stress $\sigma_\mathrm{B}$ instead of the Bludszuweit shear rate $G_\mathrm{B}$ for analogy to the power law model~\cref{eq:power_law_algebraic}. This also enables a more direct fit of the model to experimental data, which is commonly reported in terms of fluid stress.
The pore area $\poreArea$ is a function of the effective shear rate $\effShear$, which is computed from the \gls{RBC} models in \cref{sec:methods_rbc_models}.
For easier evaluation, we derive a polynomial fit, see Appendix~\ref{sec:pore_fit}:
\begin{equation}
\begin{aligned}
    \poreArea(\effShear) = 
    \begin{dcases}
        0, \quad &\effShear < G_1 \, , \\
        P_5 \left( \frac{\effShear}{G_2} \right) , \quad &G_1 \leq \effShear < G_2 \, , \\
        P_5(1), \quad &\effShear \geq G_2 \, ,
    \end{dcases}
    \\[1ex]
    P_5(x) = -13.41 x^5 + 37.31 x^4 - 48.91 x^3 + 32.39 x^2 + 0.63 x - 0.16 \, .
    \label{eq:pore_area} 
\end{aligned}
\end{equation}
\end{subequations}
Here, $G_1 = \SI{3750}{\per\second}$ and $G_2 = \SI{42000}{\per\second}$ are threshold shear rates for pore formation and lethal hemolysis, respectively. 
The parameters $h$ and $k$ are again fitted to the experimental data of \citet{dingShearInducedHemolysisSpecies2015a}. The effective shear rate $\effShear$ is computed from either the stress-based model~\eqref{eq:effShear_blud} or the strain-based model~\eqref{eq:effShear_kv}. For the strain-based model, we again use the analytical solution of the \gls{KV} model for constant fluid shear rates to compute the effective shear rate over time. The resulting parameters are summarized in \cref{tab:pore_formation_parameters}. 
\JSONParseFromFile{\stressPoreParams}{params_poreStress.json}
\JSONParseValue[rescan, store in=\stressOvineH]{\stressPoreParams}{ovine.h}
\JSONParseValue[rescan, store in=\stressBovineH]{\stressPoreParams}{bovine.h}
\JSONParseValue[rescan, store in=\stressPorcineH]{\stressPoreParams}{porcine.h}
\JSONParseValue[rescan, store in=\stressHumanH]{\stressPoreParams}{human.h}
\JSONParseFromFile{\strainPoreParams}{params_poreStrain.json}
\JSONParseValue[rescan, store in=\strainOvineH]{\strainPoreParams}{ovine.h}
\JSONParseValue[rescan, store in=\strainBovineH]{\strainPoreParams}{bovine.h}
\JSONParseValue[rescan, store in=\strainPorcineH]{\strainPoreParams}{porcine.h}
\JSONParseValue[rescan, store in=\strainHumanH]{\strainPoreParams}{human.h}

\begin{table}
    \centering
    \begin{tabular}{ll
        S[table-format=1.3e-1]
        S[table-format=1.4]
        S[table-format=1.4]}
        \toprule
        \gls{RBC} Model & Species & {$h$} & {$k$} & {Correlation coefficient ($R$)}\\
        \midrule
        stress-based & ovine & 
        \stressOvineH & 
        \JSONParseExpandableValue{\stressPoreParams}{ovine.k} & \JSONParseExpandableValue{\stressPoreParams}{ovine.r}
        \\
        & human & \stressHumanH & \JSONParseExpandableValue{\stressPoreParams}{human.k} & \JSONParseExpandableValue{\stressPoreParams}{human.r} \\
        & porcine & \stressPorcineH & \JSONParseExpandableValue{\stressPoreParams}{porcine.k} & \JSONParseExpandableValue{\stressPoreParams}{porcine.r} \\
        & bovine & \stressBovineH & \JSONParseExpandableValue{\stressPoreParams}{bovine.k} & \JSONParseExpandableValue{\stressPoreParams}{bovine.r} \\[1ex]
        strain-based & ovine & \strainOvineH & \JSONParseExpandableValue{\strainPoreParams}{ovine.k} & \JSONParseExpandableValue{\strainPoreParams}{ovine.r} \\
        & human & \strainHumanH & \JSONParseExpandableValue{\strainPoreParams}{human.k} & \JSONParseExpandableValue{\strainPoreParams}{human.r} \\
        & porcine & \strainPorcineH & \JSONParseExpandableValue{\strainPoreParams}{porcine.k} & \JSONParseExpandableValue{\strainPoreParams}{porcine.r} \\
        & bovine & \strainBovineH & \JSONParseExpandableValue{\strainPoreParams}{bovine.k} & \JSONParseExpandableValue{\strainPoreParams}{bovine.r} \\
        \bottomrule
    \end{tabular}
    \caption{Pore formation parameters for different species and \gls{RBC} models, fitted to experimental data of \citet{dingShearInducedHemolysisSpecies2015a}.}
    \label{tab:pore_formation_parameters}
\end{table}

In contrast to the power law model, the pore formation model is based on biophysical considerations~\cite{vitaleMultiscaleBiophysicalModel2014a}. Its native differential formulation in~\cref{eq:pore_formation_mddt} makes it suitable for varying shear rates without further assumptions. However, it still relies on experimental data to fit its parameters.

\subsection{Numerical Method}
\label{sec:methods_numerical_method}

Having discussed the underlying model equations for the \gls{CFD}, \gls{RBC}, and hemoglobin release models, we now present how we solve these equations. We describe boundary conditions and initial conditions first and then detail the numerical discretization and solution methods.

For the \gls{CFD}, we impose the velocity at the inlet $\Gamma_\text{in}$ and we set a no-slip boundary condition at all walls $\Gamma_\text{wall}$. At the outlet, we prescribe a zero-traction condition. As initial condition, we set zero velocity throughout the domain.

For the strain-based \gls{RBC} models, we first note that \Cref{eq:effShear_kv,eq:effShear_kvl,eq:ttlm,eq:ttlm_shear} can each be written in the following general form:
\begin{equation}
    \mDDt{\qof} = \mathbf{F}(\qof, \grad\vel) \, ,
    \label{eq:general_model}
\end{equation}
for a quantity of interest $\qof$ and right-hand side $\mathbf{F}$ depending on the specific model. There are two fundamentally different interpretations of this equation: the Lagrangian and Eulerian description.

The Lagrangian description relies on tracking individual \glspl{RBC} and evaluating the model along each trajectory. 
For a more detailed discussion of this approach, see~\cite{dirkesEulerianFormulationTensorbased2024,taskinEvaluationEulerianLagrangian2012b}. All models discussed here are available in Lagrangian form in the open-source code HemTracer (\url{https://github.com/nicodirkes/hemtracer}). 

In this work, we will focus on the Eulerian description. This description employs a continuum approach, identifying the quantity of interest as a field $\qof(\x, t)$ over the entire domain. The general equation~\eqref{eq:general_model} then becomes a \pgls{PDE} in the form of an advection equation:
\begin{equation}
    \mDDt{\qof} = \matDer{\qof} = \mathbf{F}(\qof, \grad\vel) \, ,
\end{equation}
with a source term $\mathbf{F}$. 
For the \gls{RBC} models, we prescribe undeformed cells as initial condition and steady deformation at the inlet as boundary condition. Concretely, the \gls{KV} model~\eqref{eq:effShear_kv} becomes:
\begin{alignat}{3}
\matDer{\effShear} &= \frac{1}{\charTime}\big[\bludShear - \effShear\big] \, , &\qquad& \x \in \Omega \, , &\quad& t > 0 \, , \\
\effShear &= 0 \, , &\qquad& \x \in \Omega \, , &\quad& t = 0 \, , \\
\effShear &= \bludShear \, , &\qquad& \x \in \Gamma_\text{in} \, , &\quad& t > 0 \, .
\end{alignat}

The \gls{KVL} model~\eqref{eq:effShear_kvl} is transformed accordingly. Due to the singularity at $\effShear = 0$, we set a small offset in the initial condition:
\begin{alignat}{3}
\matDer{\logEffShear} &= \bludShear e^{-\logEffShear/\charTime} - 1 \, , &\qquad& \x \in \Omega \, , &\quad& t > 0 \, , \\
\logEffShear &= -2 \, , &\qquad& \x \in \Omega \,, &\quad& t = 0 \, , \\
\logEffShear &= \charTime e^{\bludShear} \, , &\qquad& \x \in \Gamma_\text{in} \, , &\quad& t > 0 \, .
\end{alignat}

For the \gls{TTLM}~\eqref{eq:ttlm}, we employ the analytical steady state solution~\eqref{eq:steady_eigenvalues} to compute the transformed eigenvalues at the inlet:
\begin{alignat}{3}
    \matDer{\logLambI{i}} &= F_i^\mathrm{TTLM}(\logLamb; \grad\vel) \, , \qquad &&\x \in \Omega \, , &\quad& t > 0 \, , \quad i \in \{1, 3\} \, , \\
    \logLambI{i}(\x, 0) &= -2 \, , &&\x \in \Omega \, , &\quad& t = 0 \, , \\
    \logLambI{i}(\x, t) &= \logLambI{i}^\mathrm{steady}(\bludShear) \, , &&\x \in \Gamma_\text{in} \, , &\quad& t > 0 \, , 
\end{alignat}
\begin{equation}
    \eigval_1 = 1 + e^{\logLambI{1}} \, , \qquad
    \eigval_2 = \frac{1}{\eigval_1 \eigval_3} \, , \qquad
    \eigval_3 = \frac{1}{1 + e^{\logLambI{3}}} \, .
\end{equation}
The formulation for the shear-only \gls{TTLM}~\eqref{eq:ttlm_shear} is analogous.

For the hemoglobin release models, we employ the volume integral approach proposed by~\citet{garonFastThreedimensionalNumerical2004a}. This approach assumes zero hemolysis at the inlet and exploits the divergence theorem. This way, the boundary integral in \cref{eq:mih} can be computed directly from a volume integral of the hemoglobin release source term over the domain $\Omega$, avoiding the need to solve an additional transport equation for $IH$. Furthermore, it allows for more instantaneous feedback on global hemolysis in unsteady simulations, which otherwise require propagating $IH$ concentration up to the outlet.
For the power law model~\eqref{eq:power_law_algebraic}, this volume integral becomes:
\begin{equation}
    MIH(t) = \frac{10^6}{Q} 
        \left[ 
            \int_\Omega 
            A^{1/\beta} \effStress^{\alpha/\beta}(\x, t) \, \mathrm{d}\x
        \right]^\beta \, ,
    \qquad
    \effStress(\x, t) = \mu \effShear(\x, t) \, ,
    \label{eq:mih_power}
\end{equation}
For the pore formation model~\eqref{eq:pore_formation}, the volume integral becomes:
\begin{equation}
    MIH(t) = \frac{10^6}{Q}
        \int_\Omega h \sigma_\mathrm{B}^k(\x, t) \poreArea(\effShear(\x, t)) \, \mathrm{d}\x \, ,
    \qquad
    \sigma_\mathrm{B}(\x, t) = \mu G_\mathrm{B}(\x, t) \, ,
    \label{eq:mih_pore}
\end{equation}
with the piecewise polynomial $\poreArea$ from \cref{eq:pore_area}.
For unsteady simulations, we average $MIH(t)$ over time to obtain a global value $MIH$.

All problems are solved using our in-house multiphysics stabilized finite element code~\cite{dirkesEulerianFormulationTensorbased2024}. We use piecewise linear FEM as space discretization and an implicit second-order BDF2 time-stepping scheme.
We treat rotating machinery using the \gls{MRF} approach~\cite{pauliStabilizedFiniteElement2015b}. We employ \gls{VMS} stabilization, which separates the solution into resolved and unresolved scales. The application of this method to unsteady turbulent flows can be interpreted as a form of \gls{LES}~\cite{bazilevsVariationalMultiscaleResidualbased2007a}.

\section{Results}
\label{sec:results}

We present the application of these models to two benchmark cases defined by the \gls{FDA}: the \gls{FDA} blood pump and the \gls{FDA} nozzle. 

\subsection{FDA Blood Pump}
\label{sec:results_fda_pump}
 
\gls{CFD} simulations of the \gls{FDA} blood pump have previously been performed by \citet{hasslerFiniteelementFormulationAdvection2020a}. In the same work, the results have been validated against experimental \gls{PIV} data. We therefore use the same computational mesh and the same time-averaged flow field for hemolysis modeling in the present work. 

\begin{figure}
    \centering
    \begin{subfigure}{0.48\textwidth}
        \includegraphics[width=\textwidth]{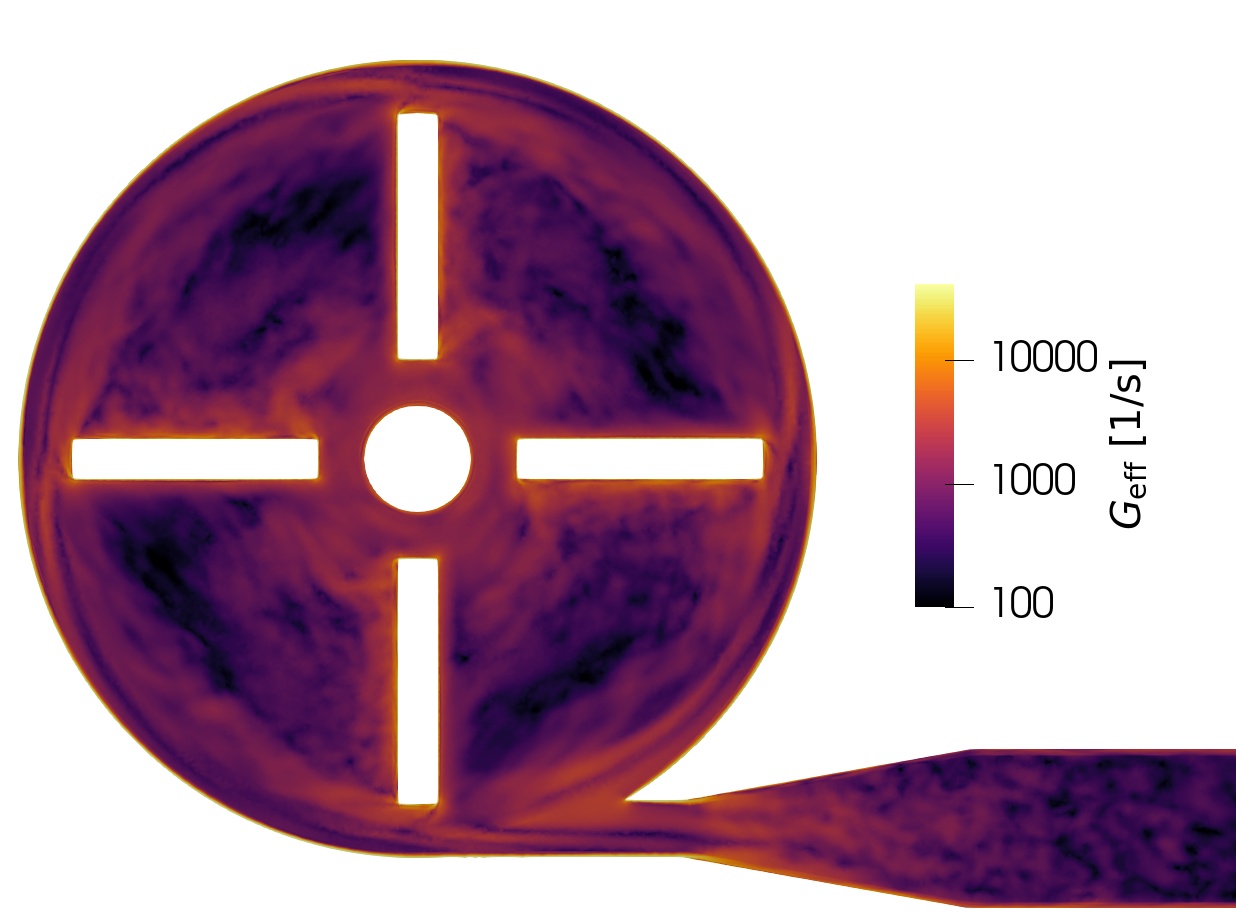}
        \caption{Bludszuweit effective shear rate~\eqref{eq:effShear_blud}}
        \label{fig:fdaPump_rbc_stressBlud}
    \end{subfigure}
    \begin{subfigure}{0.48\textwidth}
        \includegraphics[width=\textwidth]{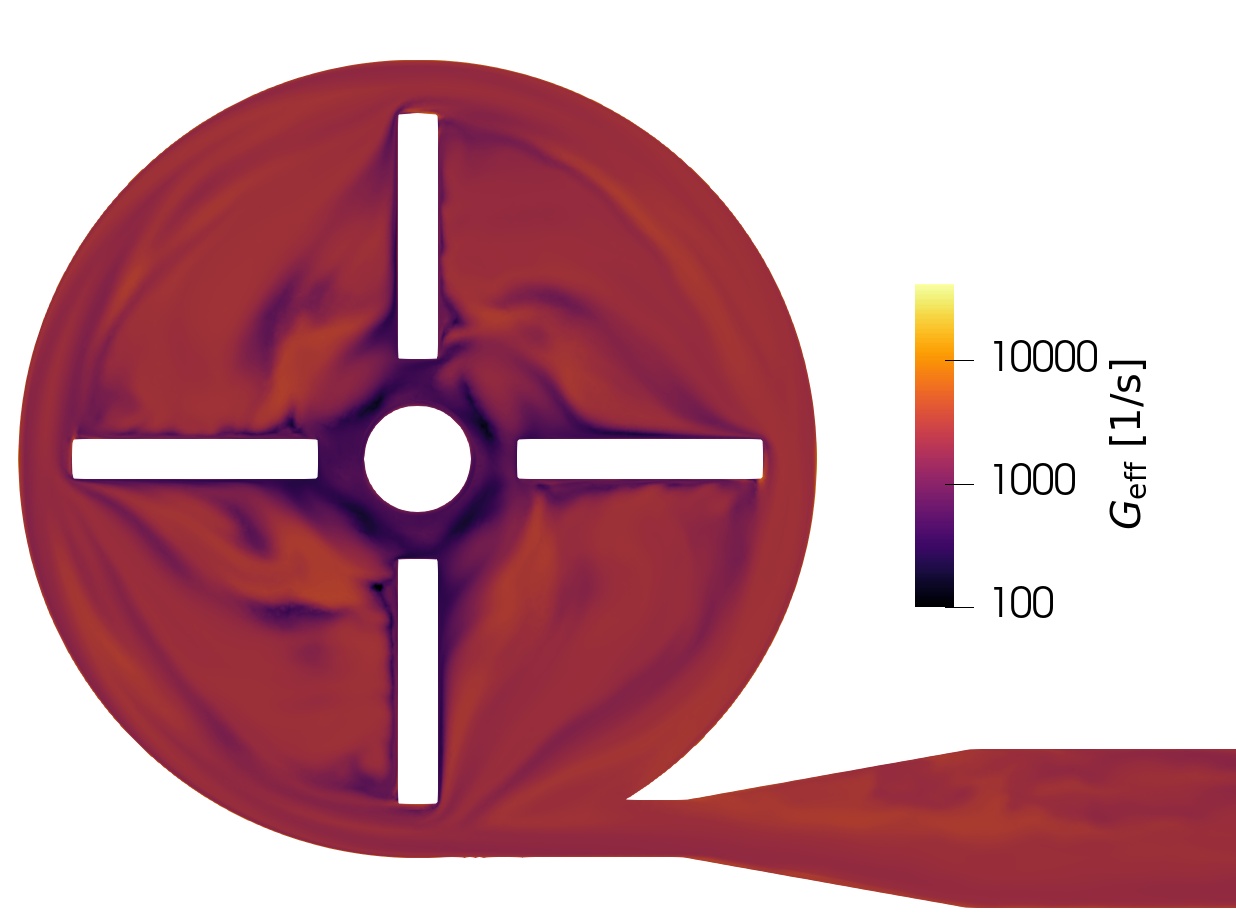}
        \caption{\gls{KV} effective shear rate~\eqref{eq:effShear_kv}}
        \label{fig:fdaPump_rbc_kv}
    \end{subfigure}
    \begin{subfigure}{0.48\textwidth}
        \includegraphics[width=\textwidth]{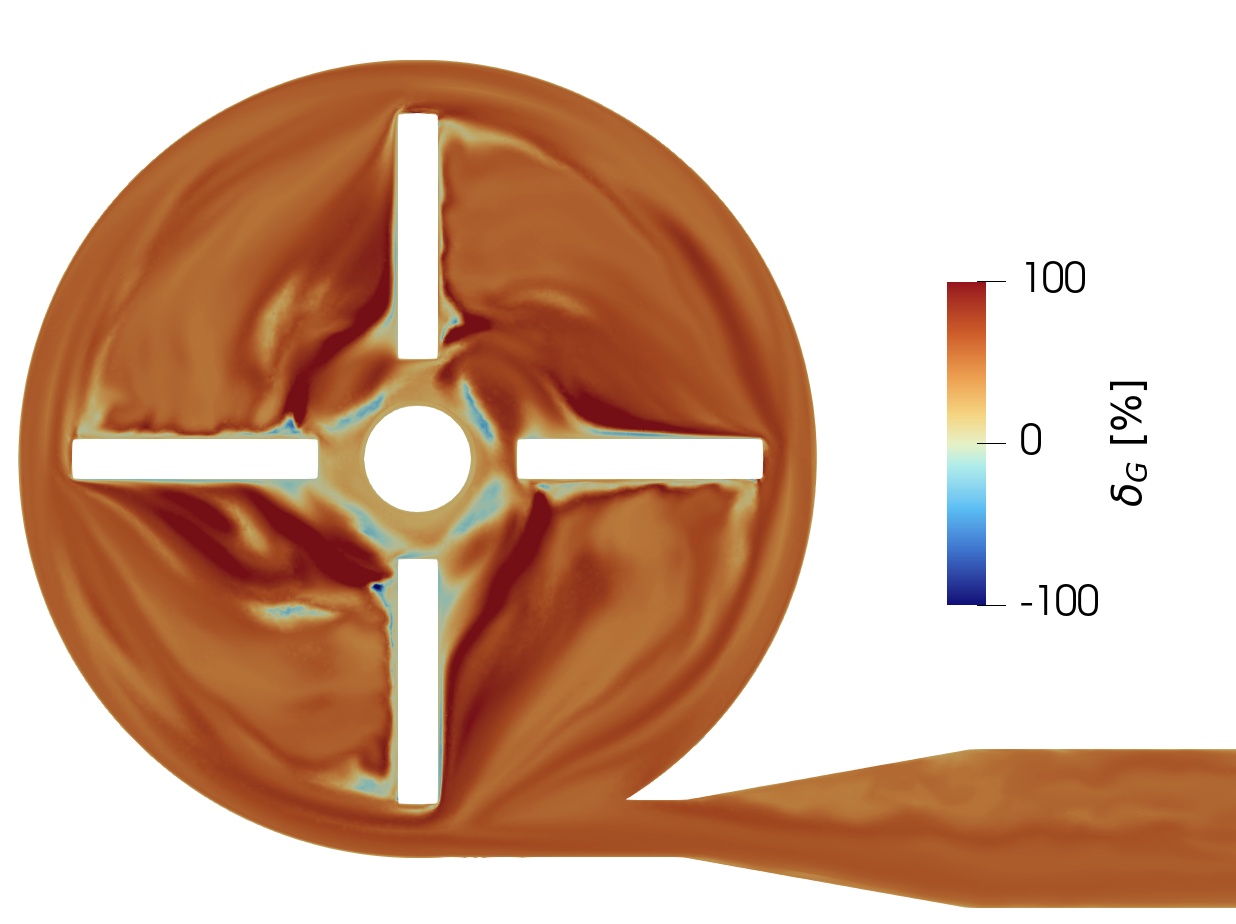}
        \caption{Difference between \gls{KV} and \gls{KVL}~\eqref{eq:deltaG_kv_kvl}}
        \label{fig:fdaPump_rbc_kv_log}
    \end{subfigure}
    \hfill
    \begin{subfigure}{0.48\textwidth}
        \includegraphics[width=\textwidth]{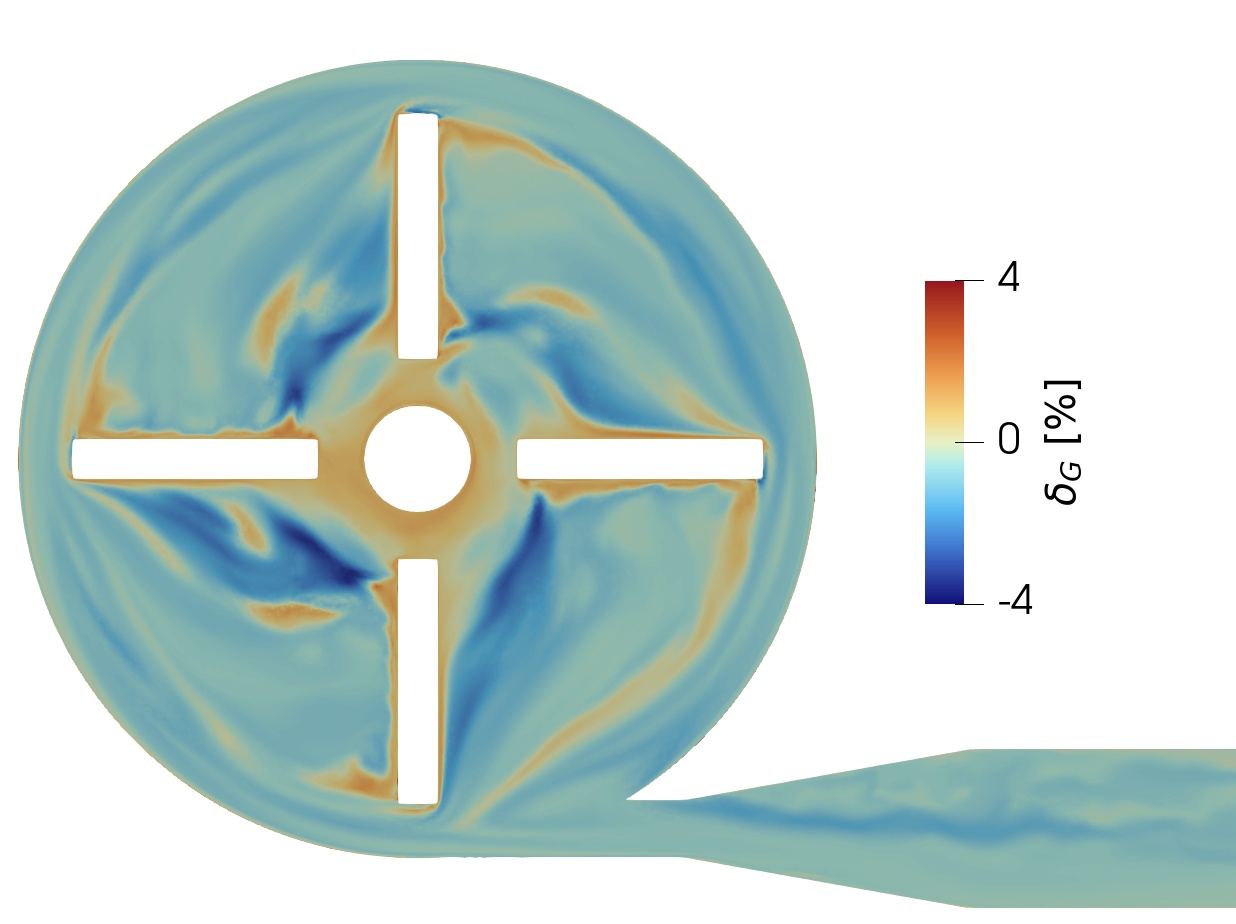}
        \caption{Difference between \gls{KVL} and \gls{TTLM}~\eqref{eq:deltaG_kvl_ttlm}}
        \label{fig:fdaPump_rbc_ttm}
    \end{subfigure}
    \begin{subfigure}{0.48\textwidth}
        \includegraphics[width=\textwidth]{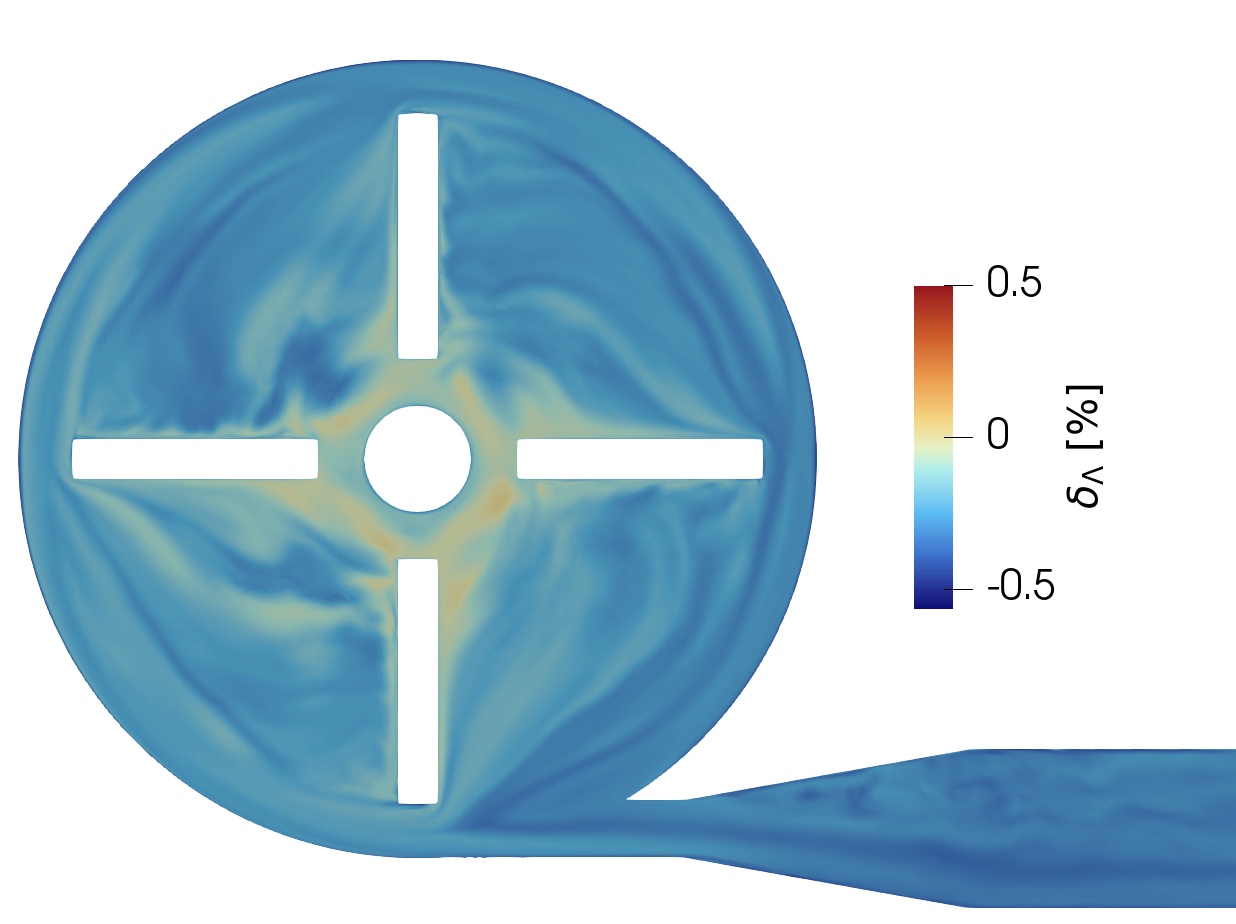}
        \caption{Orientation effect on deformation $\delta_\Lambda$~\eqref{eq:deltaL_deltaF}}
        \label{fig:fdaPump_rbc_deltaL}
    \end{subfigure}
    \hfill
    \begin{subfigure}{0.48\textwidth}
        \includegraphics[width=\textwidth]{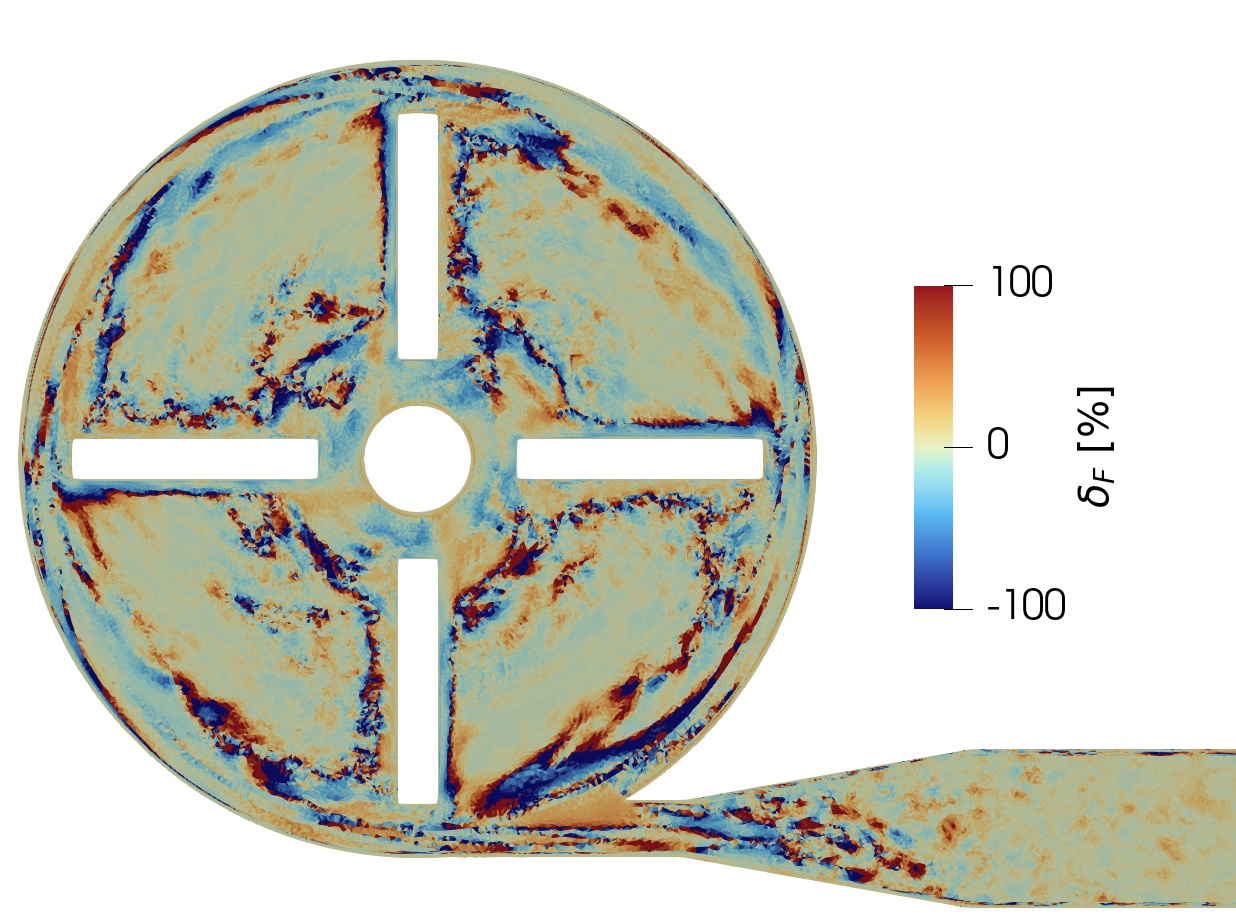}
        \caption{Orientation effect on source term $\delta_F$~\eqref{eq:deltaL_deltaF}}
        \label{fig:fdaPump_rbc_deltaF}
    \end{subfigure}
    \caption{\gls{RBC} model results for the \gls{FDA} blood pump at operating condition 4, evaluated on a plane intersecting the impeller at $z = \SI{0.7}{\centi\metre}$.}
    \label{fig:fdaPump_rbc_models}
\end{figure}

First, the flow field is used to determine $\effShear$ using the three \gls{RBC} models presented in \cref{sec:methods_rbc_models}.
We use a logarithmic scale due to the large range of effective shear rates present in the pump. The stress-based model in \cref{fig:fdaPump_rbc_stressBlud} predicts the highest effective shear rates overall, with peak values exceeding \SI{95000}{\per\second} close to the walls and to the impeller blades. In comparison, the strain-based models predict lower effective shear rates at the wall, with maximum values of approximately \SI{15000}{\per\second}. In the region in between the impeller blades, the strain-based models predict higher effective shear rates than the stress-based model due to the effect of the viscoelastic membrane relaxation. Qualitatively, the effective shear rates predicted by the three strain-based models from \cref{eq:effShear_kv,eq:effShear_kvl,eq:ttlm} are very similar. For this reason, we only show the effective shear rates from the regular \gls{KV} model in \cref{fig:fdaPump_rbc_kv} and the relative differences between the models in \cref{fig:fdaPump_rbc_kv_log,fig:fdaPump_rbc_ttm}.
Quantitatively, the predictions from the regular \gls{KV} model in \cref{fig:fdaPump_rbc_kv} are between $50\%$ and $100\%$ higher than those from the \gls{KVL} model in \cref{fig:fdaPump_rbc_kv_log}. This suggests that numerical diffusion introduced by the logarithmic formulation significantly influences the effective shear rate predictions. In contrast, \cref{fig:fdaPump_rbc_ttm} shows that the two strain-based models in logarithmic formulation yield very similar results, with differences of less than $5\%$.
Finally, the effect of \gls{RBC} orientation is analyzed in \cref{fig:fdaPump_rbc_deltaL,fig:fdaPump_rbc_deltaF}. The effective shear rates differ by less than $1\%$ between the tank-treading model and its shear-only version. However, the effect on the source term is more pronounced, with differences of up to $100\%$, as shown in \cref{fig:fdaPump_rbc_deltaF}.

\begin{figure}
    \centering
    \begin{subfigure}{0.9\textwidth}
        \includegraphics{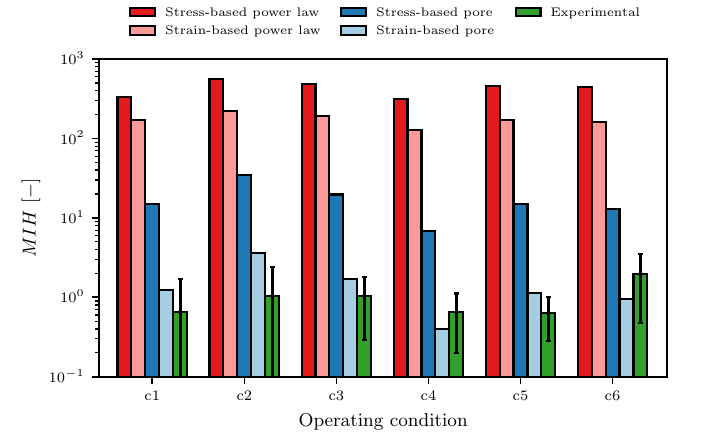}
        \caption{Stress-based (Bludszuweit) and strain-based (Kelvin-Voigt) \gls{RBC} models with power law and pore models against experiments.}
        \label{fig:fdaPump_mih_stressStrain}
    \end{subfigure}
    \begin{subfigure}{0.9\textwidth}
        \includegraphics{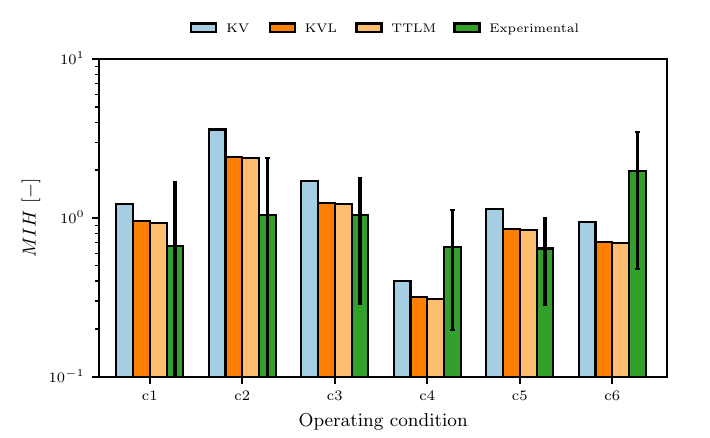}
        \caption{Different forms of strain-based \gls{RBC} models using the pore formation model for hemoglobin release.}
        \label{fig:fdaPump_mih_kv_ttm}
    \end{subfigure}
    \caption{\Gls{MIH} predictions for the \gls{FDA} pump using different \gls{RBC} models and different models for the hemoglobin release compared to experimental data from \citet{malinauskasFDABenchmarkMedical}.}
    \label{fig:fdaPump_mih}
\end{figure}

Using the effective shear rates from the \gls{RBC} models, we compute the \gls{MIH} using the volume integral approach outlined in \cref{sec:methods_numerical_method} In order to compare to the experimental data of \citet{malinauskasFDABenchmarkMedical}, which was obtained from porcine blood, we use the respective porcine parameters from~\cref{tab:pore_formation_parameters,tab:power_law_parameters}. The results for all six operating conditions of the pump are shown in \cref{fig:fdaPump_mih}. The stress-based Bludszuweit approach with the power law model overpredicts the experimental values by almost three orders of magnitude. Using the same hemoglobin model with the strain-based \gls{KV} model instead leads to $50\%$ smaller hemolysis predictions, but still significantly overpredicts the experimental values. The combination of the Bludszuweit \gls{RBC} model with the pore formation model yields lower hemolysis predictions, overpredicting the experimental values by one and a half orders of magnitude. Only the combination of the strain-based \gls{KV} model with the pore formation model leads to hemolysis predictions that are within the standard deviation of the experimental data for conditions 1, 3, 4, and 6. For conditions 2 and 5, the predictions slightly overestimate the experimental values. As \cref{fig:fdaPump_mih_kv_ttm} demonstrates, all strain-based \gls{RBC} models yield similar results. The \gls{KVL} and \gls{TTLM} models predict slightly lower values than the regular \gls{KV} model, as a result of the lower effective shear rates shown in \cref{fig:fdaPump_rbc_kv_log}. The predictions of these two models are thus within a standard deviation for all conditions. Between the \gls{KVL} and \gls{TTLM} model, there are practically no differences in the predicted \gls{MIH} values, confirming the observations made in \cref{fig:fdaPump_rbc_ttm}.

\subsection{FDA Nozzle}
\label{sec:results_fda_nozzle}

As a second benchmark case, we consider the \gls{FDA} nozzle~\cite{hariharanMultilaboratoryParticleImage2011a}. Similar to \citet{mantegazzaExaminingUniversalityHemolysis2023b}, we focus on the sudden contraction orientation at a Reynolds number of $Re = 6500$, as this is the only configuration with both PIV data and hemolysis data. We employ a block-structured computational mesh with approximately 2.4 million hexahedral elements. In the throat region, the wall distance is approximately \SI{12.5}{\micro\metre}, which is within the Kolmogorov length scale of \SI{20}{\micro\metre} reported by \citet{mantegazzaExaminingUniversalityHemolysis2023b}. We set a time step size of \SI{1e-5}{\second} and simulate until a statistically stationary quasi-steady state is reached. From there, we compute the unsteady flow solution using the \gls{LES} approach outlined in \cref{sec:methods_numerical_method} for a total time of \SI{20}{\milli\second}. For the simulations, we employ a viscosity of \SI{4.24e-3}{\pascal\second}, a density of \SI{1040}{\kilo\gram\per\metre\cubed}, and a volume flow rate of \SI{6.77e-5}{\metre\cubed\per\second}. This corresponds to the conditions in the hemolysis experiments with bovine blood, as reported by \citet{herbertsonMultilaboratoryStudyFlowInduced2015}. 

\begin{figure}
    \centering
    \begin{subfigure}{\textwidth}
        \centering
        \includegraphics[width=\textwidth]{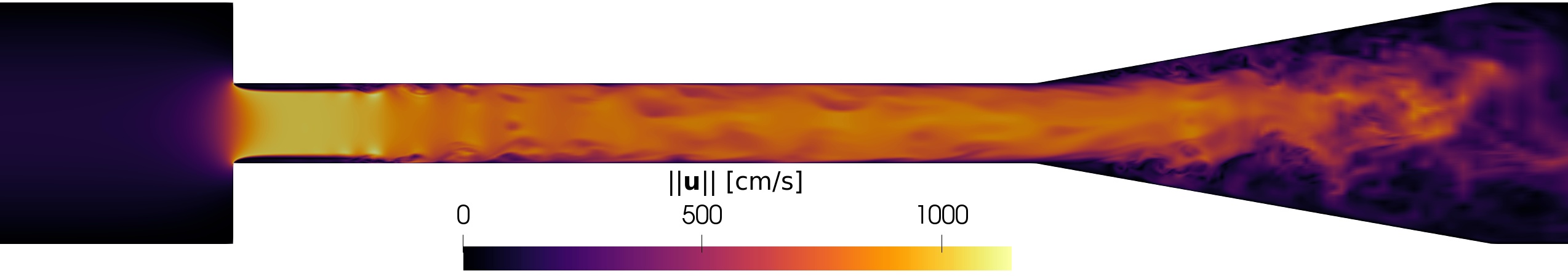}
        \caption{Instantaneous velocity magnitude from \gls{LES} solution.}
        \label{fig:fdaNozzle_velocity_instant}
    \end{subfigure}
    \begin{subfigure}{\textwidth}
        \centering
        \includegraphics[width=\textwidth]{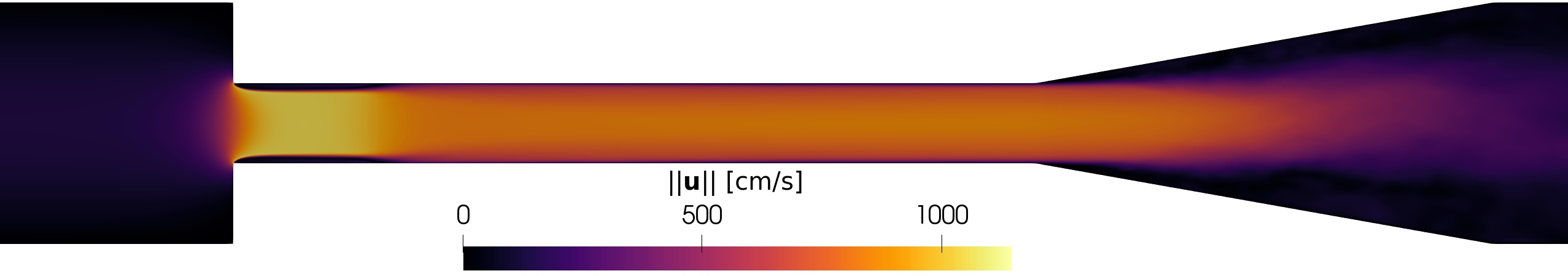}
        \caption{Time-averaged velocity magnitude from \gls{RANS} solution.}
        \label{fig:fdaNozzle_velocity_avg}
    \end{subfigure}
    \begin{subfigure}{0.48\textwidth}
        \includegraphics{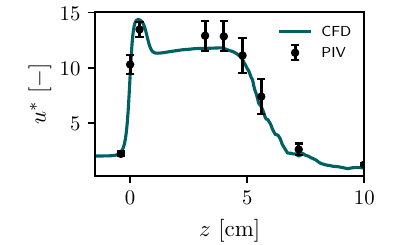}
        \caption{Non-dimensional RANS axial velocity along centerline compared to \gls{PIV} data.}
        \label{fig:fdaNozzle_velocity_centerline}
    \end{subfigure}
    \hfill
    \begin{subfigure}{0.48\textwidth}
        \includegraphics{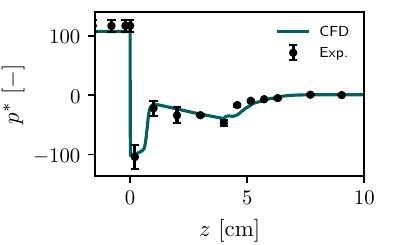}
        \caption{Non-dimensional \gls{RANS} wall pressure compared to experimental data.}
        \label{fig:fdaNozzle_pressure_centerline}
    \end{subfigure}
    \begin{subfigure}{0.48\textwidth}
        \includegraphics{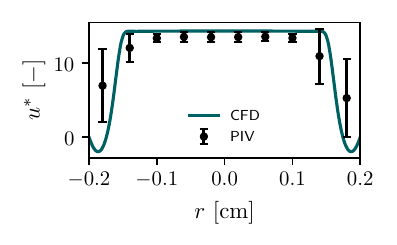}
        \caption{Non-dimensional \gls{RANS} velocity profile at $z = 0.4\,\si{cm}$ (throat region) compared to \gls{PIV} data.}
        \label{fig:fdaNozzle_velocity_profile_1}
    \end{subfigure}
    \hfill
    \begin{subfigure}{0.48\textwidth}
        \includegraphics{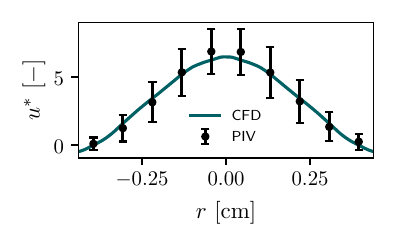}
        \caption{Non-dimensional \gls{RANS} velocity profile at $z = 5.6\,\si{cm}$ (diffuser region) compared to \gls{PIV} data.}
        \label{fig:fdaNozzle_velocity_profile_2}
    \end{subfigure}
    \caption{Flow field validation for the FDA nozzle in contraction orientation at $Re = 6500$ using data from~\citet{hariharanMultilaboratoryParticleImage2011a}.}
    \label{fig:fdaNozzle_velocity}
\end{figure}

In order to validate the flow solution with \gls{PIV} data, we average the flow field in time over \SI{20}{\milli\second}. Since the \gls{PIV} data was obtained by \citet{hariharanMultilaboratoryParticleImage2011a} with water at a higher flow rate, we non-dimensionalize the velocity by the mean inlet velocity $\bar{u}$. Similarly, we non-dimensionalize the pressure by the corresponding dynamic pressure $\rho \bar{u}^2 / 2$. Since the \gls{PIV} measurements were performed at the same Reynolds number $Re = 6500$ as the simulations, dynamic scaling guarantees similarity. We compute the wall pressure by averaging the pressure values at the wall over the circumference at each axial position. 
\Cref{fig:fdaNozzle_velocity} shows the \gls{CFD} results. As \cref{fig:fdaNozzle_velocity_instant} visualizes, the flow in the throat region is highly unsteady, with vortices forming and shedding downstream of the sudden expansion. \Cref{fig:fdaNozzle_velocity_avg} shows the corresponding time-averaged velocity field. This field is used as basis for the experimental validation. \Cref{fig:fdaNozzle_velocity_centerline,fig:fdaNozzle_pressure_centerline,fig:fdaNozzle_velocity_profile_1,fig:fdaNozzle_velocity_profile_2} compare the \gls{CFD} results to the data from \citet{hariharanMultilaboratoryParticleImage2011a}. The error bars indicate the standard deviation of the experiments. The axial direction $z$ is defined in direction of the flow, with $z=\SI{0}{\centi\metre}$ at the sudden contraction. Overall, the velocity predictions along the centerline in \cref{fig:fdaNozzle_velocity_centerline} show very good agreement with the experimental data, with errors within the standard deviation. The predicted wall pressure in \cref{fig:fdaNozzle_pressure_centerline} is mostly within the experimental standard deviation, with some errors outside the standard deviation in the expansion region. The velocity profile in \cref{fig:fdaNozzle_velocity_profile_1} at $z=\SI{0.4}{\centi\metre}$ is in the throat region immediately downstream of the contraction. The comparison shows less recirculation in the experiments compared to the \gls{RANS} results, leading to deviations in the velocity profile close to the wall. Similar deviations have been observed by \citet{mantegazzaExaminingUniversalityHemolysis2023b}. Since the standard deviation is large in this region, there may be strong transient effects present in the experiments that are not fully captured by the time-averaged \gls{RANS} solution. In contrast, the velocity profile in \cref{fig:fdaNozzle_velocity_profile_2} at $z=\SI{5.6}{\centi\metre}$ in the diffuser region shows very good agreement with the experimental data throughout the profile. 

\def\widthFigRBC{\textwidth}
\begin{figure}
    \centering
    \begin{subfigure}{\textwidth}
        \centering
        \includegraphics[width=\widthFigRBC]{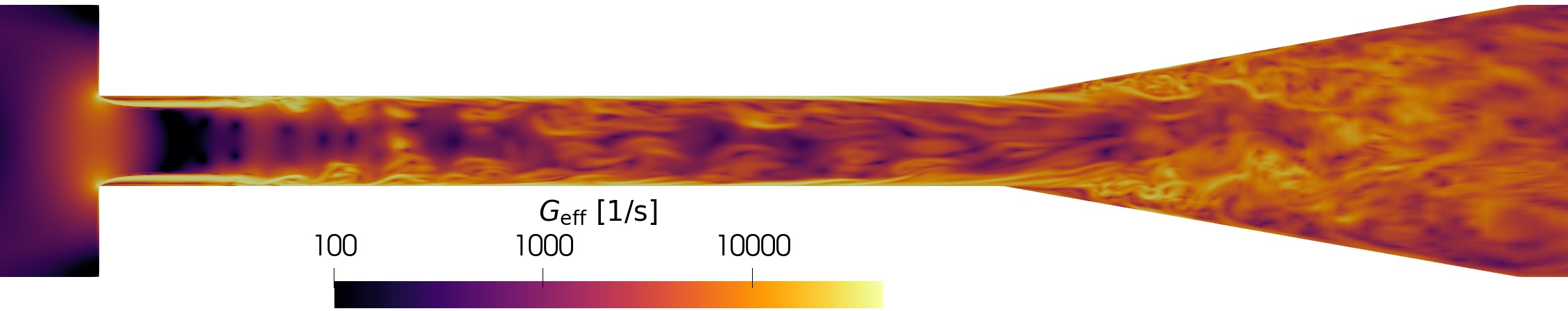}
        \caption{Effective shear rate from Bludszuweit model, see \cref{eq:effShear_blud}}
        \label{fig:fdaNozzle_rbc_stressBlud}
    \end{subfigure}
    \begin{subfigure}{\textwidth}
        \centering
        \includegraphics[width=\widthFigRBC]{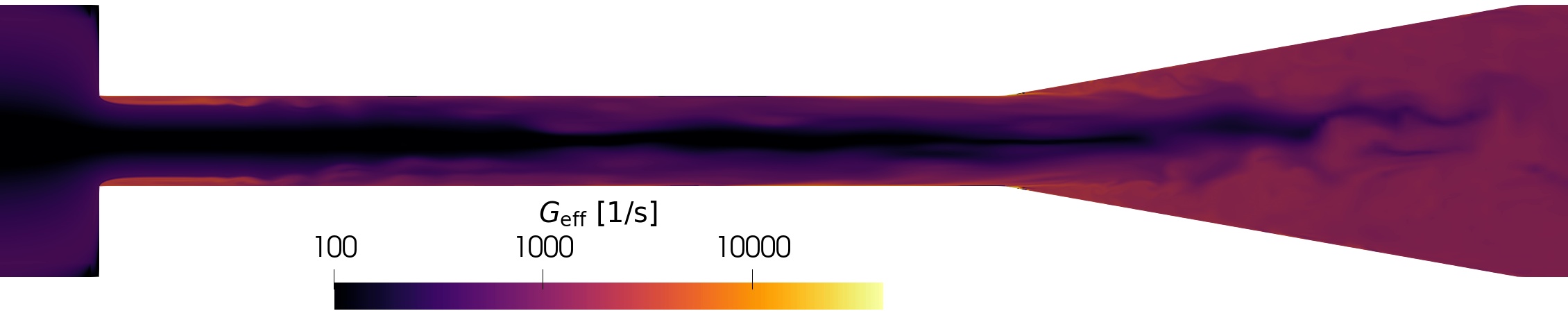}
        \caption{Effective shear rate from \gls{KV} model, see \cref{eq:effShear_kv}}
        \label{fig:fdaNozzle_rbc_kv}
    \end{subfigure}
    \begin{subfigure}{\textwidth}
        \centering
        \includegraphics[width=\widthFigRBC]{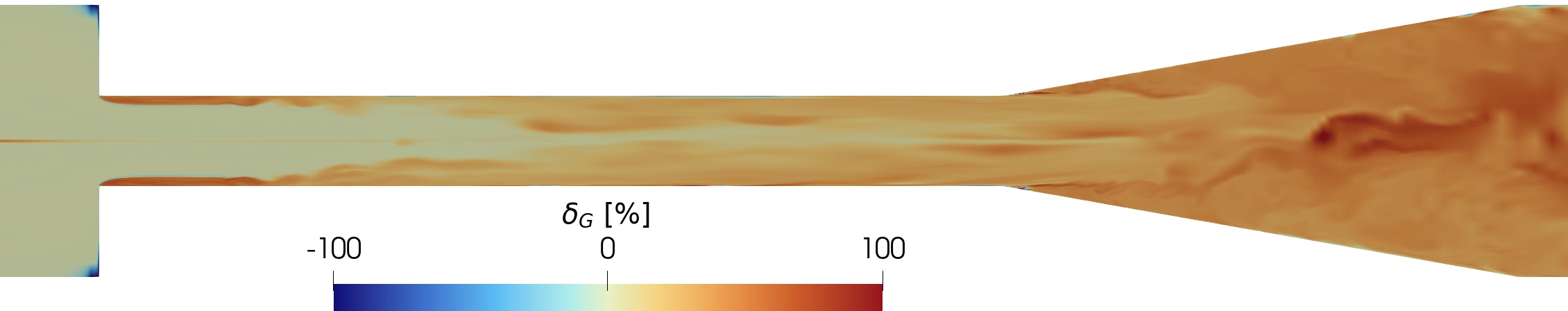}
        \caption{Difference between \gls{KV} and \gls{KVL}~$\delta_{G}^{\mathrm{KV};\mathrm{KVL}}$, see \cref{eq:deltaG_kv_kvl}}
        \label{fig:fdaNozzle_rbc_kv_log}
    \end{subfigure}
    \begin{subfigure}{\textwidth}
        \centering
        \includegraphics[width=\widthFigRBC]{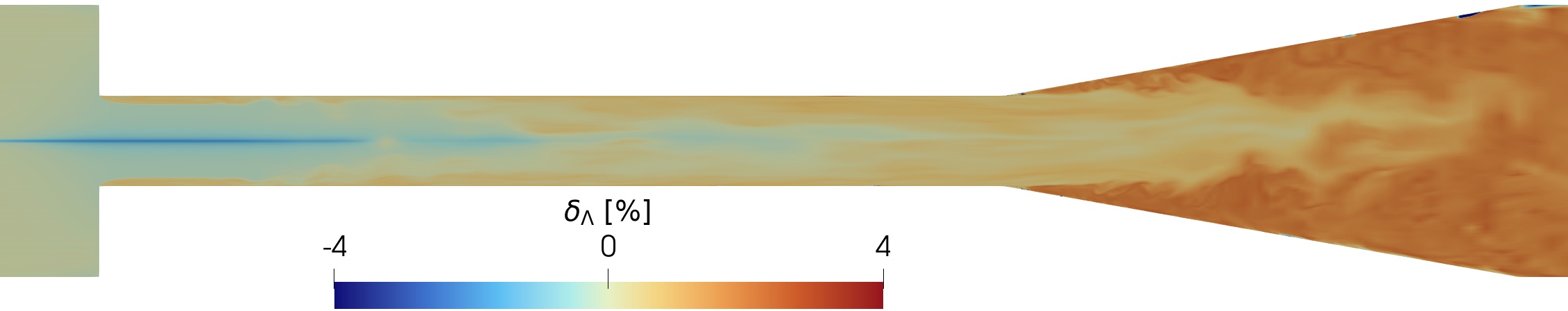}
        \caption{Orientation effect on deformation, see \cref{eq:deltaL_deltaF}}
        \label{fig:fdaNozzle_rbc_deltaL}
    \end{subfigure}
    \begin{subfigure}{\textwidth}
        \centering
        \includegraphics[width=\widthFigRBC]{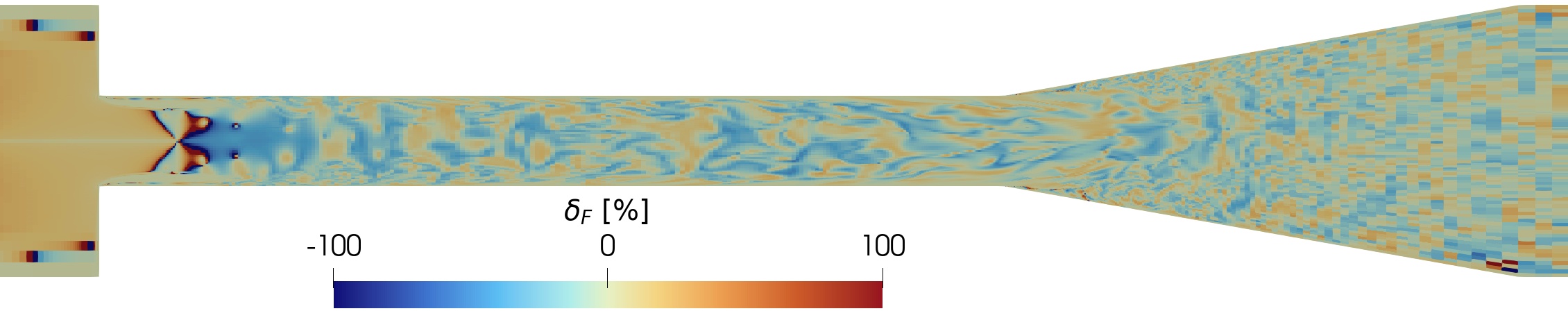}
        \caption{Orientation effect on source term, see \cref{eq:deltaL_deltaF}}
        \label{fig:fdaNozzle_rbc_deltaF}
    \end{subfigure}
    \caption{\gls{RBC} model results for the \gls{FDA} nozzle in contraction configuration at $Re = 6500$, evaluated on a central axial plane.}
    \label{fig:fdaNozzle_rbc_models}
\end{figure}

Next, we discuss the results of the various \gls{RBC} models. We apply these models to the unsteady \gls{LES} flow field. The fields visualized in \cref{fig:fdaNozzle_rbc_models} show the effective shear rates at one particular instant. The effective shear rates using the stress-based Bludszuweit model in \cref{fig:fdaNozzle_rbc_stressBlud} again exhibit the largest values of all models, with peak values in the nozzle up to \SI{3e5}{\per\second}. In comparison, the strain-based \gls{KV} model in \cref{fig:fdaNozzle_rbc_kv} predicts lower effective shear rates, with peak values in the nozzle of approximately \SI{1e4}{\per\second}. Qualitatively, the \gls{KV} model predicts a larger coherent region of elevated $\effShear$ in the recirculation zone immediately downstream of the contraction. The relative difference between the \gls{KV} and \gls{KVL} models in \cref{fig:fdaNozzle_rbc_kv_log} shows that the \gls{KV} model again predicts up to $100\%$ higher effective shear rates compared to the \gls{KVL} model. The difference between the \gls{TTLM} and \gls{KVL} models is again much smaller and omitted here for brevity. Finally, the effect of the three-dimensional \gls{RBC} orientation is shown in \cref{fig:fdaNozzle_rbc_deltaL,fig:fdaNozzle_rbc_deltaF}. We observe the same trends as in the blood pump case in \cref{fig:fdaPump_rbc_deltaL,fig:fdaPump_rbc_deltaF}: the effect on the source term is significant, but the resulting deformations are very similar.

\begin{figure}
    \centering
    \begin{subfigure}{0.48\textwidth}
        \includegraphics{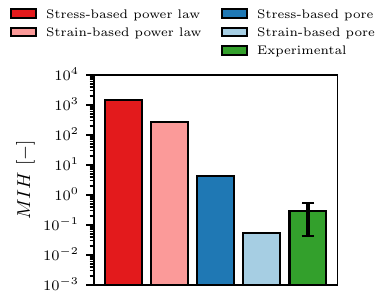}
        \caption{Stress-based (Bludszuweit) and strain-based (\gls{KV}) \gls{RBC} models with power law and pore models for hemoglobin release.}
        \label{fig:fdaNozzle_mih_stressStrain}
    \end{subfigure}
    \hfill
    \begin{subfigure}{0.48\textwidth}
        \includegraphics{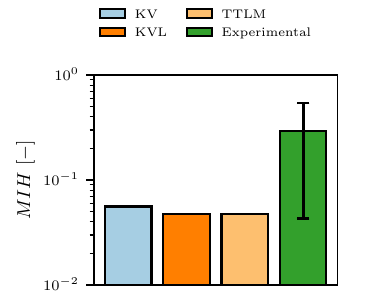}
        \caption{Different versions of strain-based \gls{RBC} models using the pore formation model for hemoglobin release.}
        \label{fig:fdaNozzle_mih_kv_ttm}
    \end{subfigure}
    \caption{\Gls{MIH} predictions for the \gls{FDA} nozzle in contraction configuration at $Re = 6500$, using different \gls{RBC} models and hemoglobin release models, compared to experimental data from \citet{herbertsonMultilaboratoryStudyFlowInduced2015}.}
    \label{fig:fdaNozzle_mih}
\end{figure}

Lastly, we compute the \gls{MIH} using the volume integral approach from \cref{sec:methods_numerical_method}. Since the hemolysis experiments by \citet{herbertsonMultilaboratoryStudyFlowInduced2015} were performed with bovine blood, we use the respective bovine parameters from~\cref{tab:pore_formation_parameters,tab:power_law_parameters}. The results are shown in \cref{fig:fdaNozzle_mih}. Similar to the blood pump case in \cref{fig:fdaPump_mih}, the power law models overpredict the experimental values by three to four orders of magnitude. The pore formation models achieve better predictions overall. In particular, the strain-based \gls{RBC} model in combination with the pore formation model yield predictions that are within the experimental standard deviation. The difference between the \gls{KV}, \gls{KVL}, and \gls{TTLM} models is relatively small, with the \gls{KVL} and \gls{TTLM} models predicting slightly lower hemolysis than the regular \gls{KV} model. In contrast, the stress-based Bludszuweit \gls{RBC} in combination with the pore formation model overestimates the experimental values by approximately one order of magnitude. The power law model again leads to significant overpredictions between three and four orders of magnitude. Such overprediction is also observed in the results presented by \citet{mantegazzaExaminingUniversalityHemolysis2023b}.

\section{Discussion}
\label{sec:discussion}

Our results show that the strain-based pore formation model consistently outperforms the other combinations of \gls{RBC} models and hemoglobin release models in both benchmark cases. It is able to accurately predict hemolysis over a wide range of operating conditions using a single set of model parameters fitted to literature data. This demonstrates the potential of the strain-based pore formation model as a universal tool for computational hemolysis prediction in complex blood flows. Its predictive capability is particularly notable given the simplicity of the model, which only requires solving a single linear advection equation for the effective shear rate and evaluating a volume integral to compute hemolysis. The model thus accounts for the underlying viscoelastic \gls{RBC} deformation on a phenomenological level, without the need for explicit modeling of shape and orientation dynamics or interactions between individual cells.

Our results further confirm that Eulerian power law models tend to overpredict hemolysis, see also \cite{ponnaluriResultsInterlaboratoryComputational2023b,taskinEvaluationEulerianLagrangian2012b,faghihPracticalImplicationsErroneous2023}. Since the original form of the power law~\eqref{eq:power_law_algebraic} is written for constant stress in simple experimental conditions, any application to complex flows incurs various types of errors. For example, the linearization required for non-constant stress histories leads to numerical errors. In addition, the power law is purely empirical in nature. So while it may fit experimental data well in simple shear flow, there is no guarantee that it will generalize to more complex flow conditions. In contrast, the pore formation model is based on biophysical principles of membrane poration, making it more robust to variations in flow conditions. Its differential formulation~\eqref{eq:pore_formation_mddt} allows it to naturally account for non-constant stress or strain histories without the need for linearization.

The current default approach for computational hemolysis prediction is the stress-based power law model (see the employed models in \citet{ponnaluriResultsInterlaboratoryComputational2023b}). While the absolute hemolysis predictions of this approach are often inaccurate, it can still be useful for relative comparisons~\cite{mantegazzaExaminingUniversalityHemolysis2023b}. It is therefore often argued that the stress-based power law model is sufficient for distinguishing between iterative design variations of medical devices. However, as shown by \citet{lommelExperimentalInvestigationApplicability}, stress-based models are not able to accurately capture the effect of short-term exposure to high shear stresses, which are common in medical devices. If the exposure time to these peaks or the magnitude of the peaks varies between different device designs, stress-based models may yield misleading results even for relative comparisons. In contrast, strain-based models are able to account for the viscoelastic response of \glspl{RBC} to transient shear stresses, leading to more robust predictions in such scenarios.
For these reasons, we find that strain-based models are generally preferable over stress-based models for computational hemolysis prediction. 

On the different forms of strain-based \gls{RBC} models, namely \gls{KV}, \gls{KVL}, and \gls{TTLM}, our results indicate that the choice between them has only a minor impact on hemolysis predictions. In fact, we find larger variations between the logarithmic and regular formulation of the \acrfull{KV} model than between \gls{KVL} and \gls{TTLM}. While the \gls{KV} and \gls{KVL} formulations are analytically equivalent, the piecewise linear finite element discretization of the logarithmic form introduces numerical diffusion, leading to lower effective shear rates and thus lower hemolysis predictions. With increasing mesh resolution, this difference is expected to decrease. The \gls{KVL} and \gls{TTLM} models, on the other hand, differ mainly in their ability to account for the difference between shear and elongational flows. The difference between these models seems to be negligible for both benchmarks considered here. We employ the shear-only \gls{TTLM} to analyze the reasons for this. We find that there is a significant influence of elongational versus shear stresses on the source terms. In accordance with our earlier analysis~\cite{dirkesSignificanceFlowVorticity2025}, there are significant coherent regions in the flow where the incorporation of these three-dimensional effects causes the source term to either increase or decrease substantially compared to the shear-only assumption. However, the effect on the resulting \gls{RBC} deformation is small. We postulate that this is due to the short exposure times to either elevated or reduced source terms. Due to the characteristic deformation timescale of \SI{200}{\milli\second}, the \glspl{RBC} do not have enough time to respond to these local variations before they are advected into a different flow region with different stress characteristics. Overall, these effects seem to cancel out, leading to similar predictions for both \gls{KVL} and \gls{TTLM}. There is thus little practical benefit in using the more complex \gls{TTLM} over the \gls{KV} or \gls{KVL} models for typical medical device flows.

Next, we discuss the limitations of our modeling approach. First, the strain-based pore formation model does not explicitly resolve single cells or their microstructural behavior. In particular, it does not account for interactions between \glspl{RBC} or detailed membrane mechanics.
Instead, it models the average response of a population of \glspl{RBC} to the local flow conditions using models informed by biophysical principles. While this phenomenological approach captures the values and trends of hemolysis well in the benchmark cases considered here, further validation against experimental data in real medical devices is needed to establish its generalizability.
Second, these results do not include any coupling between \gls{RBC} deformation and the surrounding fluid flow, assuming Newtonian fluid behavior. In reality, deformed \glspl{RBC} affect the rheological properties of blood, which in turn affect the flow field. In a primitive approach, this could be accounted for by using shear-thinning viscosity models in the \gls{CFD} simulation. A more rigorous approach would involve two-way coupling between \gls{RBC} deformation and fluid flow.
Third, our models assume constant hematocrit and do not account for variations in \gls{RBC} concentration or aggregation effects, which can influence hemoglobin release. Incorporating these effects would require more complex multiphase flow models~\cite{melkaNumericalInvestigationMultiphase2019}. 
Fourth, the strain-based \gls{RBC} models presented here only account for the stresses induced by the resolved flow field. In turbulent flows, additional stresses arise from velocity fluctuations at scales smaller than the grid resolution. 
For stress-based models, the inclusion of such stresses on the macroscopic scale has been studied using Reynolds shear stress~\cite{goubergritsTurbulenceBloodDamage2016} or energy dissipation rate~\cite{wuRepresentationEffectiveStress2019}. For strain-based models, it is still an open question how incorporate turbulence effects, especially due to the large discrepancy between the timescales of turbulent fluctuations and \gls{RBC} deformation.

In the future, we plan to extend our modeling approach in several directions. First, we intend to apply our workflow to real ventricular assist devices in order to further validate its predictive capability in practical applications. 
Second, we are working on incorporating two-way coupling between \gls{RBC} deformation and fluid flow in the context of viscoelastic fluid models~\cite{bodnarSimulationThreeDimensionalFlow2011}. %
Third, we aim to perform uncertainty quantification for our hemolysis predictions and thus obtain estimates for the confidence intervals of our model outputs~\cite{blumUncertaintyAwareHemolysisModeling2025a}. This will identify the magnitude of uncertainties propagating from the experimental data used for model calibration~\cite{blumQuantifyingExperimentalVariability2025}. Fourth, we want to obtain highly resolved turbulence data in order to study the effects of turbulent fluctuations on \gls{RBC} deformation and hemolysis and develop suitable modeling approaches to account for these effects on a phenomenological level.

Overall, we have discovered a promising new approach for computational hemolysis prediction, consisting of a \acrfull{KV}-based constitutive model for \gls{RBC} deformation combined with a pore formation model for hemoglobin release. This approach consistently outperforms existing methods across two benchmark cases, and is simple to implement into existing \gls{CFD} workflows. It can be applied to any converged \gls{CFD} solution directly in the Eulerian frame, enabling detailed local analysis of \gls{RBC} deformation as well as accurate global predictions of hemolysis. The improvements in accuracy are attributed to the biophysical basis of the \gls{KV} model, which captures the viscoelastic deformation behavior of \gls{RBC} membranes, as well as the pore formation model, which describes hemoglobin release through membrane pores.

\backmatter
\section*{Statements and Declarations}

\subsection*{Competing Interests}
The authors have no relevant financial or non-financial interests to disclose.

\subsection*{Funding}
This work was funded by the Deutsche Forschungsgemeinschaft (DFG, German Research Foundation) through grant 333849990/GRK2379 (IRTG Modern Inverse Problems). 
The authors gratefully acknowledge the computing time provided to them at the NHR Centers NHR4CES at TU Darmstadt (project number p0020502) and RWTH Aachen University (project number p0024828). This is funded by the Federal Ministry of Research, Technology and Space, and the state governments participating on the basis of the resolutions of the GWK for national high performance computing at universities (\url{www.nhr-verein.de/unsere-partner}).

\subsection*{Acknowledgements}
The authors would like to thank Dr. Zhongjun Jon Wu for providing us with the raw experimental data from \citet{dingShearInducedHemolysisSpecies2015a} for the parameter fitting in \cref{tab:power_law_parameters,tab:pore_formation_parameters}.

\subsection*{CRediT authorship contribution statement}
\textbf{Nico Dirkes}: Conceptualization, Methodology, Software, Validation, Formal analysis, Investigation, Data Curation, Writing - Original Draft, Visualization, Software. \textbf{Marek Behr}: Conceptualization, Writing - Review \& Editing, Supervision, Project administration, Funding acquisition, Software.

\appendix
\numberwithin{equation}{section}
\begin{appendices}
\section{Evaluation of the Pore Area}
\label{sec:pore_fit}

The pore model derived by~\citet{vitaleMultiscaleBiophysicalModel2014a} defines the total pore area $\poreArea$ per \gls{RBC} as a function of the surface area strain $\surfStrain$ of the \gls{RBC} membrane:
\begin{equation}
    \surfStrain = \frac{A_\mathrm{m} - A_0}{A_0} \, ,
    \label{eq:surface_strain}
\end{equation}
where $A_\mathrm{m}$ is the current membrane surface area and $A_0$ is the surface area of an undeformed \gls{RBC}. This expression limits the pore model's applicability to \gls{RBC} models that explicitly resolve the \gls{RBC} shape, such as the \gls{TTM}, see \cref{eq:ttm}. In this section, we derive the pore area as a function of the effective shear rate $\effShear$ instead. This allows for the application of the pore model to any \gls{RBC} model that provides the effective shear rate. This section is split into two parts: First, we reiterate the model equations that relate the pore area to the surface area strain according to \citet{vitaleMultiscaleBiophysicalModel2014a}. Second, we describe how to compute the surface area strain for the models discussed in \cref{sec:methods_rbc_models} and derive appropriate polynomial fits for easy evaluation. All parameters used in this section are summarized in \cref{tab:pore_fit_params}.
\begin{table}
    \centering
    \begin{tabular}{crll}
        \toprule
    Symbol & Value & Unit & Description \\
        \midrule
    $A_0$ & 135 & \si{\micro\metre\squared} & Unstretched \gls{RBC} surface area \\
    $\sigma^\prime$ & 3.2 & \si{\milli\joule\per\metre\squared} & Apparent hydrophobicity \\
    $\xi$ & 8 &  & Lipid tail rigidity \\
    $h_t$ & 2.5 & \si{\nano\metre} & Length of phospholipid tails \\
    $r_\mathrm{l}$ & \num{9.5e-2} & \si{\nano\metre} & Radius of phospholipid heads \\
    $N_1$ & \num{7.913e-3} & \si{\per\micro\metre\squared} & Pore density at $\surfStrain_1$, computed from \cref{eq:pore_N1} \\
    $N_2$ & 108.97 & \si{\per\micro\metre\squared} & Pore density at $\surfStrain_2$, computed from \cref{eq:pore_N2} \\
    $A_2^\star$ & 5.5 & $\%$ & Relative pore area at $\surfStrain_2$ \\
    $\surfStrain_1$ & 0.16 & $\%$ & Threshold surface strain for pore formation \\
    $\surfStrain_2$ & 6.0 & $\%$ & Surface strain at lethal hemolysis \\
    $f_1$ & 5.0 & \si{\second^{-1}} & Relaxation parameter in RBC model \\
    $f_2$ & \num{4.2298e-4}     &  & Steady shear RBC deformation parameter \\
    $\effShear^{(1)}$ & 3750 & \si{\per\second} & Threshold effective shear rate, computed from \cref{eq:G1_G2} \\
    $\effShear^{(2)}$ & 42000 & \si{\per\second} & Lethal effective shear rate, computed from \cref{eq:G1_G2} \\
    $a_0$ & \num{-0.15929974}   & \si{\micro\metre\squared} & Polynomial fit coefficient, computed from \cref{eq:polynomial_fit} \\
    $a_1$ & \num{0.63319421}    & \si{\micro\metre\squared} & Polynomial fit coefficient, computed from \cref{eq:polynomial_fit} \\
    $a_2$ & \num{32.39208945}   & \si{\micro\metre\squared} & Polynomial fit coefficient, computed from \cref{eq:polynomial_fit} \\
    $a_3$ & \num{-48.91487332}  & \si{\micro\metre\squared} & Polynomial fit coefficient, computed from \cref{eq:polynomial_fit} \\
    $a_4$ & \num{37.31248604}   & \si{\micro\metre\squared} & Polynomial fit coefficient, computed from \cref{eq:polynomial_fit} \\
    $a_5$ & \num{-13.41339481}  & \si{\micro\metre\squared} & Polynomial fit coefficient, computed from \cref{eq:polynomial_fit} \\
        \bottomrule
    \end{tabular}
    \caption{Parameters used for the evaluation of the pore area $\poreArea$. If not stated otherwise, the values are taken from \citet{vitaleMultiscaleBiophysicalModel2014a}.}
    \label{tab:pore_fit_params}
\end{table}

We start by considering the normalized membrane energy, expressed as a fourth order polynomial of the pore radius $r_0$ and the pore density $N$:
\begin{align}
    \hat{W}_N(r_0) = A N^2 r_0^4 - B N r_0^2 + C N r_0 \, , \label{eq:W_r} \\
    A(\surfStrain) = 2 \pi^2 \left[\sigma^\prime \xi - \sigma_0^\prime(\surfStrain)\right] \, , \qquad
    B = \sigma_0^\prime(\surfStrain) \pi \, , \qquad
    C = 2 \pi \gamma(\surfStrain) \, . \label{eq:W_ABC}
\end{align}
Note that the expression for $A$ in \cref{eq:W_ABC} differs from the one given by \citet{vitaleMultiscaleBiophysicalModel2014a}. With the original definition, we were not able to reproduce the values of $N_1$ and $N_2$ reported in their work. With the present definition, we obtain similar values (comp. \cref{tab:pore_fit_params}).
The surface tension of the intact membrane $\sigma_0^\prime$ is computed as:
\begin{equation}
    \sigma_0^\prime(\surfStrain) = 2 \sigma^\prime \xi \left( 1 - \frac{1}{(1 + \surfStrain)^2} \right) \, .
    \label{eq:sigma0_prime}
\end{equation}
The pore edge energy $\gamma$ is computed as:
\begin{equation}
    \gamma(\surfStrain) = 2 \left(  h_t - \xi r_\mathrm{l} \surfStrain \right) \sigma^\prime \, .
    \label{eq:gamma}
\end{equation}
The critical pore density $N_\mathrm{c}$ at which pores start to form is given by:
\begin{equation}
    N_\mathrm{c}(\surfStrain) = \frac{8 B(\surfStrain)^3}{27 A(\surfStrain) C(\surfStrain)^2} \, .
    \label{eq:pore_Nc}
\end{equation}
As in \cite{vitaleMultiscaleBiophysicalModel2014a}, we compute $N_1$ from \cref{eq:pore_Nc} as
\begin{equation}
    N_1 = N_\mathrm{c}(\surfStrain_1) \, ,
    \label{eq:pore_N1}
\end{equation}
and $N_2$ as:
\begin{equation}
    r_2 = \frac{- \pi C(\surfStrain_2)}{4 A(\surfStrain_2) A_2^\star - 2 \pi B(\surfStrain_2)} \, , \qquad
    N_2 = \frac{2 B(\surfStrain_2) r_2 - C(\surfStrain_2)}{4 A(\surfStrain_2) r_2^3}
    \, .
    \label{eq:pore_N2}
\end{equation}
The pore density is modeled as a function of the surface strain $\surfStrain$ by an exponential interpolation between $N_1$ and $N_2$:
\begin{equation}
    N(\surfStrain) = N_1 \exp \left( \frac{\surfStrain - \surfStrain_1}{\surfStrain_1 - \surfStrain_2} \ln \left( \frac{N_1}{N_2} \right) \right) \, , \quad
    \surfStrain_1 \leq \surfStrain \leq \surfStrain_2 \, .
    \label{eq:pore_N}
\end{equation}
The stable pore radius $\poreRadius$ is now computed as the local minimum of the normalized membrane energy $\hat{W}_N$ with respect to $r_0$.
With the closure~\eqref{eq:pore_N}, analysis of \cref{eq:W_r} shows that a local minimum at some $r_0^\star > 0$ exists for all $\surfStrain \in [\surfStrain_1, \surfStrain_2]$. The position of this minimum can then be calculated analytically: 
\begin{equation}
    r_0^\star (\surfStrain) = 
    2 \sqrt{\frac{B(\surfStrain)}{6 A(\surfStrain) N(\surfStrain)}} \cos \left( \frac{1}{3} \arccos \left( \frac{-3 C(\surfStrain)}{4 B(\surfStrain)} \sqrt{\frac{6 A(\surfStrain) N(\surfStrain)}{B(\surfStrain)}} \right) \right) \, .
\end{equation}
Below $\surfStrain_1$, no pores are forming. Above $\surfStrain_2$, the membrane is considered to be fully ruptured. The pore radius $\poreRadius$ is thus given by:
\begin{equation}
    \poreRadius = 
    \begin{dcases}
    0 \, , & \surfStrain < \surfStrain_1 \, , \\
    r_0^\star (\surfStrain) \, , & \surfStrain_1 \leq \surfStrain \leq \surfStrain_2 \, , \\
    r_0^\star (\surfStrain_2) \, , & \surfStrain > \surfStrain_2 \, .
    \end{dcases}
    \label{eq:pore_radius}
\end{equation}
Finally, the total pore area for a single \gls{RBC} is given by:
\begin{equation}
    \poreArea = \Phi_{\surfStrain}(\surfStrain) \coloneqq \pi (1 + \surfStrain) A_0 N(\surfStrain) \poreRadius^2(\surfStrain)  \, ,
    \label{eq:pore_rP}
\end{equation}
with the expressions from~\cref{eq:pore_radius,eq:pore_N}. The function $\Phi_{\surfStrain}$ thus defines the pore area in terms of the surface strain $\surfStrain$. This has been the model as introduced by \citet{vitaleMultiscaleBiophysicalModel2014a}.

Next, we derive a function 
\begin{equation}
    \surfStrain = \Psi_G(\effShear) \, ,
    \label{eq:surface_strain_effShear}
\end{equation}
which will allow us to evaluate the surface strain $\surfStrain$ for any of the models in \cref{sec:methods_rbc_models}. First, we determine the surface area of an arbitrary ellipsoid with semi-axes $(a,b,c)$. For this purpose, we use the convergent series by \citet{kellerSurfaceAreaEllipsoid1979}:
\begin{align}
    A_\mathrm{ell} (a,b,c; \ n) =
    2 \pi b c
    +
    \frac{4\pi a b}{n}
    \sum_{j=1}^{n/2}
    \frac{1-\tau_j}{\sqrt{1-\left(\frac{c}{a}\right)^2-\tau_j}}\,
    \arcsin
    \sqrt{\frac{1-\left(\frac{c}{a}\right)^2-\tau_j}{\,1-\tau_j\,}} \, , \\
    t_j = \cos\left(\frac{(2j-1)\,\pi}{2n}\right), 
    \qquad
    \tau_j = \left(1-\bigl(\tfrac{c}{b}\bigr)^2\right)\, t_j^2,
    \qquad
    n\ \text{even},
\end{align}
which converges to the true surface area of an ellipsoid as $n \to \infty$.
In our computations, we use $n=60$. 
For the \gls{TTM}, we thus compute the surface area using the semi-axes obtained from \cref{eq:ttm_deformation}:
\begin{equation}
    A_\mathrm{TTM}(\Lamb) = A_\mathrm{ell} (\sqrt{\eigval_1}, \sqrt{\eigval_2}, \sqrt{\eigval_3} ; \ 60) \, .
    \label{eq:surface_area_ttm}
\end{equation}
At lethal hemolysis, real \glspl{RBC} have $1.06$ times the surface area of their resting state $A_0$~\cite{aroraTensorBasedMeasureEstimating2004}. The ellipsoidal representation in the \gls{TTM} leads to excess surface area in their resting state (sphere). As a result, lethal hemolysis in the ellipsoidal representation occurs at $1.4 \times 1.06$ times the original surface area $A_\mathrm{s}$~\cite{aroraTensorBasedMeasureEstimating2004}. Assuming a linear relationship between the ellipsoidal surface strain in the \gls{TTM} model and the surface strain of the \gls{RBC}, we can thus relate the two as follows:
\begin{equation}
    \surfStrain^{(2)} = \frac{1.06 A_0 - A_0}{A_0} \, , \quad
    \surfStrain^{(2)}_\mathrm{TTM} = \frac{1.4 \cdot 1.06 A_\mathrm{s} - A_\mathrm{s}}{A_\mathrm{s}} \, , \quad
    \surfStrain = \frac{\surfStrain^{(2)}}{\surfStrain^{(2)}_\mathrm{TTM}} \surfStrain_\mathrm{TTM}  = \frac{15}{121} \surfStrain_\mathrm{TTM} \, .
    \label{eq:surface_strain_scaling} 
\end{equation}
From \cref{eq:surface_area_ttm,eq:surface_strain_scaling}, we compute the surface strain for any deformation~$\Lamb$  as:
\begin{equation}
    \surfStrain = \Psi_{\Lamb} (\Lamb) \coloneqq \frac{A_\mathrm{TTM}(\Lamb) - A_\mathrm{s}}{A_\mathrm{s}} \cdot \frac{15}{121} \, , \qquad
    A_\mathrm{s} = 4 \pi (\lambda_1 \lambda_2 \lambda_3)^{1/3} \, ,
    \label{eq:surface_strain_lambda}
\end{equation}
where $A_\mathrm{s}$ is the surface area of a sphere with the same volume as the ellipsoid defined by $\Lamb$.

\begin{figure}
    \centering
    \includegraphics{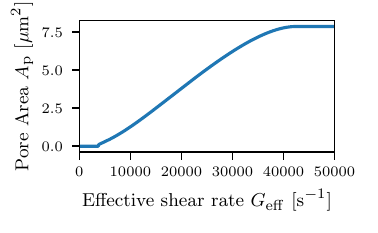}
    \caption{Pore area $\poreArea$ as a function of the effective shear rate $\effShear$.}
    \label{fig:Ap_Geff}
\end{figure}

Finally, we need to find a way to determine the equivalent ellipsoidal shape $\Lamb$ from the effective shear rate $\effShear$ for the other \gls{RBC} models discussed in \cref{sec:methods_rbc_models}. For this purpose, we employ the steady state shear deformation values derived by \citet{aroraTensorBasedMeasureEstimating2004}:
\begin{equation}
\begin{aligned}
    \eigval_1^\mathrm{steady}(G) &=
    \left( \frac{f_1^2}{f_1^2 + f_2^2 G^2} \right)^{1/3} 
    \frac{f_1^2 + f_2^2 G^2 + f_2 G \sqrt{f_1^2 + f_2^2 G^2}}{f_1^2} \, , \\
    \eigval_2^\mathrm{steady}(G) &= 
    \left( \frac{f_1^2}{f_1^2 + f_2^2 G^2} \right)^{1/3}
    \, , \\
    \eigval_3^\mathrm{steady}(G) &=
    \left( \frac{f_1^2}{f_1^2 + f_2^2 G^2} \right)^{1/3} 
     \frac{f_1^2 + f_2^2 G^2 - f_2 G \sqrt{f_1^2 + f_2^2 G^2}}{f_1^2} \, .
\end{aligned}\label{eq:steady_eigenvalues}
\end{equation}
These give the steady state deformation of an ellipsoidal \gls{RBC} in simple shear flow with shear rate $G$. By evaluating these expressions at the effective shear rate $\effShear$ provided by the \gls{RBC} models, we obtain an estimate for the instantaneous equivalent ellipsoidal shape:
\begin{equation}
    \Lamb_\mathrm{eff}(\effShear) =
    \diag \left(
    \eigval_1^\mathrm{steady}(\effShear) \, ,
    \eigval_2^\mathrm{steady}(\effShear) \, ,
    \eigval_3^\mathrm{steady}(\effShear)
    \right) \, .
    \label{eq:lambda_eff}
\end{equation}
The surface strain can now be computed from \cref{eq:surface_strain_lambda,eq:lambda_eff} as
\begin{equation}
    \surfStrain = \Psi_G(\effShear) \coloneqq \Psi_{\Lamb} \left( \Lamb_\mathrm{eff}(\effShear) \right) \, .
    \label{eq:surface_strain_G}
\end{equation}
This gives us the desired relation~\eqref{eq:surface_strain_effShear} between the surface strain $\surfStrain$ and the effective shear rate $\effShear$.
We determine the threshold effective shear rate $\effShear^{(1)}$ and the lethal effective shear rate $\effShear^{(2)}$ such that
\begin{equation}
    \Psi_G \left( \effShear^{(1)} \right) = \surfStrain_1 \, , \qquad
    \Psi_G \left( \effShear^{(2)} \right) = \surfStrain_2 \, .
    \label{eq:G1_G2}
\end{equation}
In the range between $\effShear^{(1)}$ and $\effShear^{(2)}$, we can thus evaluate the total pore area as a function of the effective shear rate by combining \cref{eq:surface_strain_G,eq:pore_rP}:
\begin{equation}
    \poreArea = \Phi_G(\effShear) \coloneqq \Phi_{\surfStrain} \left( \Psi_G (\effShear) \right) = \Phi_{\surfStrain} \left( \Psi_{\Lamb} \left( \Lamb_\mathrm{eff}(\effShear) \right) \right) \, ,
    \label{eq:pore_area_shear}
\end{equation}
The resulting pore area $\poreArea$ as a function of the effective shear rate $\effShear$ is shown in \cref{fig:Ap_Geff}.
Finally, we evaluate \cref{eq:pore_area_shear} for 10000 values of $\effShear$ linearly spaced between $\effShear^{(1)}$ and $\effShear^{(2)}$ and fit a fifth-order polynomial $P_5$ to the resulting data using least squares minimization, i.e.,
\begin{equation}
    P_5(\effShear/G_2) = \sum_{i=0}^5 a_i (\effShear/G_2)^i 
    \approx \Phi_G(\effShear) \, , \qquad
    G_1 \leq \effShear \leq G_2 \, .
    \label{eq:polynomial_fit}
\end{equation}
We obtain a polynomial fit with a mean square error of \SI{3.3124e-6}{\micro\meter^4}. This yields the simplified expression for the pore area as a function of the effective shear rate:
\begin{equation}
    P_5(x) = \sum_{i=0}^5 a_i x^i \, , \qquad
    \poreArea(\effShear)
    = 
    \begin{dcases}
    0 \, , & \effShear < G_1 \, , \\
    P_5(\effShear/G_2) \, , & G_1 \leq \effShear \leq G_2 \, , \\
    P_5(1) \, , & \effShear > G_2 \, .
    \end{dcases}
\end{equation}

\end{appendices}

\bibliography{Hemoglobin_paper}


\begin{thebibliography}{57}
\ifx \bisbn   \undefined \def \bisbn  #1{ISBN #1}\fi
\ifx \binits  \undefined \def \binits#1{#1}\fi
\ifx \bauthor  \undefined \def \bauthor#1{#1}\fi
\ifx \batitle  \undefined \def \batitle#1{#1}\fi
\ifx \bjtitle  \undefined \def \bjtitle#1{#1}\fi
\ifx \bvolume  \undefined \def \bvolume#1{\textbf{#1}}\fi
\ifx \byear  \undefined \def \byear#1{#1}\fi
\ifx \bissue  \undefined \def \bissue#1{#1}\fi
\ifx \bfpage  \undefined \def \bfpage#1{#1}\fi
\ifx \blpage  \undefined \def \blpage #1{#1}\fi
\ifx \burl  \undefined \def \burl#1{\textsf{#1}}\fi
\ifx \doiurl  \undefined \def \doiurl#1{\url{https://doi.org/#1}}\fi
\ifx \betal  \undefined \def \betal{\textit{et al.}}\fi
\ifx \binstitute  \undefined \def \binstitute#1{#1}\fi
\ifx \binstitutionaled  \undefined \def \binstitutionaled#1{#1}\fi
\ifx \bctitle  \undefined \def \bctitle#1{#1}\fi
\ifx \beditor  \undefined \def \beditor#1{#1}\fi
\ifx \bpublisher  \undefined \def \bpublisher#1{#1}\fi
\ifx \bbtitle  \undefined \def \bbtitle#1{#1}\fi
\ifx \bedition  \undefined \def \bedition#1{#1}\fi
\ifx \bseriesno  \undefined \def \bseriesno#1{#1}\fi
\ifx \blocation  \undefined \def \blocation#1{#1}\fi
\ifx \bsertitle  \undefined \def \bsertitle#1{#1}\fi
\ifx \bsnm \undefined \def \bsnm#1{#1}\fi
\ifx \bsuffix \undefined \def \bsuffix#1{#1}\fi
\ifx \bparticle \undefined \def \bparticle#1{#1}\fi
\ifx \barticle \undefined \def \barticle#1{#1}\fi
\bibcommenthead
\ifx \bconfdate \undefined \def \bconfdate #1{#1}\fi
\ifx \botherref \undefined \def \botherref #1{#1}\fi
\ifx \url \undefined \def \url#1{\textsf{#1}}\fi
\ifx \bchapter \undefined \def \bchapter#1{#1}\fi
\ifx \bbook \undefined \def \bbook#1{#1}\fi
\ifx \bcomment \undefined \def \bcomment#1{#1}\fi
\ifx \oauthor \undefined \def \oauthor#1{#1}\fi
\ifx \citeauthoryear \undefined \def \citeauthoryear#1{#1}\fi
\ifx \endbibitem  \undefined \def \endbibitem {}\fi
\ifx \bconflocation  \undefined \def \bconflocation#1{#1}\fi
\ifx \arxivurl  \undefined \def \arxivurl#1{\textsf{#1}}\fi
\csname PreBibitemsHook\endcsname

\bibitem[\protect\citeauthoryear{Katz
  et~al.}{2015}]{katzMulticenterAnalysisClinical2015c}
\begin{barticle}
\bauthor{\bsnm{Katz}, \binits{J.N.}},
\bauthor{\bsnm{Jensen}, \binits{B.C.}},
\bauthor{\bsnm{Chang}, \binits{P.P.}},
\bauthor{\bsnm{Myers}, \binits{S.L.}},
\bauthor{\bsnm{Pagani}, \binits{F.D.}},
\bauthor{\bsnm{Kirklin}, \binits{J.K.}}:
\batitle{A {{Multicenter Analysis}} of {{Clinical Hemolysis}} in {{Patients
  Supported}} with {{Durable}}, {{Long-Term Left Ventricular Assist Device
  Therapy}}}.
\bjtitle{The Journal of Heart and Lung Transplantation}
\bvolume{34}(\bissue{5}),
\bfpage{701}--\blpage{709}
(\byear{2015})
\doiurl{10.1016/j.healun.2014.10.002}
\end{barticle}
\endbibitem

\bibitem[\protect\citeauthoryear{Shah
  et~al.}{2017}]{shahBleedingThrombosisAssociated2017}
\begin{barticle}
\bauthor{\bsnm{Shah}, \binits{P.}},
\bauthor{\bsnm{Tantry}, \binits{U.S.}},
\bauthor{\bsnm{Bliden}, \binits{K.P.}},
\bauthor{\bsnm{Gurbel}, \binits{P.A.}}:
\batitle{Bleeding and thrombosis associated with ventricular assist device
  therapy}.
\bjtitle{The Journal of Heart and Lung Transplantation}
\bvolume{36}(\bissue{11}),
\bfpage{1164}--\blpage{1173}
(\byear{2017})
\doiurl{10.1016/j.healun.2017.05.008}
\end{barticle}
\endbibitem

\bibitem[\protect\citeauthoryear{Omar
  et~al.}{2015}]{omarPlasmaFreeHemoglobin2015}
\begin{barticle}
\bauthor{\bsnm{Omar}, \binits{H.R.}},
\bauthor{\bsnm{Mirsaeidi}, \binits{M.}},
\bauthor{\bsnm{Socias}, \binits{S.}},
\bauthor{\bsnm{Sprenker}, \binits{C.}},
\bauthor{\bsnm{Caldeira}, \binits{C.}},
\bauthor{\bsnm{Camporesi}, \binits{E.M.}},
\bauthor{\bsnm{Mangar}, \binits{D.}}:
\batitle{Plasma {{Free Hemoglobin Is}} an {{Independent Predictor}} of
  {{Mortality}} among {{Patients}} on {{Extracorporeal Membrane Oxygenation
  Support}}}.
\bjtitle{PLOS ONE}
\bvolume{10}(\bissue{4}),
\bfpage{0124034}
(\byear{2015})
\doiurl{10.1371/journal.pone.0124034}
\end{barticle}
\endbibitem

\bibitem[\protect\citeauthoryear{Lyu
  et~al.}{2016}]{lyuPlasmaFreeHemoglobin2016}
\begin{barticle}
\bauthor{\bsnm{Lyu}, \binits{L.}},
\bauthor{\bsnm{Long}, \binits{C.}},
\bauthor{\bsnm{Hei}, \binits{F.}},
\bauthor{\bsnm{Ji}, \binits{B.}},
\bauthor{\bsnm{Liu}, \binits{J.}},
\bauthor{\bsnm{Yu}, \binits{K.}},
\bauthor{\bsnm{Chen}, \binits{L.}},
\bauthor{\bsnm{Yao}, \binits{J.}},
\bauthor{\bsnm{Hu}, \binits{Q.}},
\bauthor{\bsnm{Hu}, \binits{J.}},
\bauthor{\bsnm{Gao}, \binits{G.}}:
\batitle{Plasma {{Free Hemoglobin Is}} a {{Predictor}} of {{Acute Renal Failure
  During Adult Venous-Arterial Extracorporeal Membrane Oxygenation Support}}}.
\bjtitle{Journal of Cardiothoracic and Vascular Anesthesia}
\bvolume{30}(\bissue{4}),
\bfpage{891}--\blpage{895}
(\byear{2016})
\doiurl{10.1053/j.jvca.2016.02.011}
\end{barticle}
\endbibitem

\bibitem[\protect\citeauthoryear{Ponnaluri
  et~al.}{2023}]{ponnaluriResultsInterlaboratoryComputational2023b}
\begin{barticle}
\bauthor{\bsnm{Ponnaluri}, \binits{S.V.}},
\bauthor{\bsnm{Hariharan}, \binits{P.}},
\bauthor{\bsnm{Herbertson}, \binits{L.H.}},
\bauthor{\bsnm{Manning}, \binits{K.B.}},
\bauthor{\bsnm{Malinauskas}, \binits{R.A.}},
\bauthor{\bsnm{Craven}, \binits{B.A.}}:
\batitle{Results of the {{Interlaboratory Computational Fluid Dynamics Study}}
  of the {{FDA Benchmark Blood Pump}}}.
\bjtitle{Annals of Biomedical Engineering}
\bvolume{51}(\bissue{1}),
\bfpage{253}--\blpage{269}
(\byear{2023})
\doiurl{10.1007/s10439-022-03105-w}
\end{barticle}
\endbibitem

\bibitem[\protect\citeauthoryear{Ezzeldin
  et~al.}{2015}]{ezzeldinStrainBasedModelMechanical2015b}
\begin{barticle}
\bauthor{\bsnm{Ezzeldin}, \binits{H.M.}},
\bauthor{\bsnm{{de Tullio}}, \binits{M.D.}},
\bauthor{\bsnm{Vanella}, \binits{M.}},
\bauthor{\bsnm{Solares}, \binits{S.D.}},
\bauthor{\bsnm{Balaras}, \binits{E.}}:
\batitle{A {{Strain-Based Model}} for {{Mechanical Hemolysis Based}} on a
  {{Coarse-Grained Red Blood Cell Model}}}.
\bjtitle{Annals of Biomedical Engineering}
\bvolume{43}(\bissue{6}),
\bfpage{1398}--\blpage{1409}
(\byear{2015})
\doiurl{10.1007/s10439-015-1273-z}
\end{barticle}
\endbibitem

\bibitem[\protect\citeauthoryear{Z{\'a}vodszky
  et~al.}{2017}]{zavodszkyCellularLevelInSilico2017b}
\begin{botherref}
\oauthor{\bsnm{Z{\'a}vodszky}, \binits{G.}},
\oauthor{\bsnm{{van Rooij}}, \binits{B.}},
\oauthor{\bsnm{Azizi}, \binits{V.}},
\oauthor{\bsnm{Hoekstra}, \binits{A.}}:
Cellular {{Level In-Silico Modeling}} of {{Blood Rheology}} with an {{Improved
  Material Model}} for {{Red Blood Cells}}.
Frontiers in Physiology
\textbf{8}
(2017)
\doiurl{10.3389/fphys.2017.00563}
\end{botherref}
\endbibitem

\bibitem[\protect\citeauthoryear{Mendez
  et~al.}{2014}]{mendezUnstructuredSolverSimulations2014}
\begin{barticle}
\bauthor{\bsnm{Mendez}, \binits{S.}},
\bauthor{\bsnm{Gibaud}, \binits{E.}},
\bauthor{\bsnm{Nicoud}, \binits{F.}}:
\batitle{An unstructured solver for simulations of deformable particles in
  flows at arbitrary {{Reynolds}} numbers}.
\bjtitle{Journal of Computational Physics}
\bvolume{256},
\bfpage{465}--\blpage{483}
(\byear{2014})
\doiurl{10.1016/j.jcp.2013.08.061}
\end{barticle}
\endbibitem

\bibitem[\protect\citeauthoryear{Fedosov
  et~al.}{2010}]{fedosovMultiscaleRedBlood2010b}
\begin{barticle}
\bauthor{\bsnm{Fedosov}, \binits{D.A.}},
\bauthor{\bsnm{Caswell}, \binits{B.}},
\bauthor{\bsnm{Karniadakis}, \binits{G.E.}}:
\batitle{A {{Multiscale Red Blood Cell Model}} with {{Accurate Mechanics}},
  {{Rheology}}, and {{Dynamics}}}.
\bjtitle{Biophysical Journal}
\bvolume{98}(\bissue{10}),
\bfpage{2215}--\blpage{2225}
(\byear{2010})
\doiurl{10.1016/j.bpj.2010.02.002}
\end{barticle}
\endbibitem

\bibitem[\protect\citeauthoryear{Guglietta
  et~al.}{2021}]{gugliettaLoadingRelaxationDynamics2021b}
\begin{barticle}
\bauthor{\bsnm{Guglietta}, \binits{F.}},
\bauthor{\bsnm{Behr}, \binits{M.}},
\bauthor{\bsnm{Falcucci}, \binits{G.}},
\bauthor{\bsnm{Sbragaglia}, \binits{M.}}:
\batitle{Loading and {{Relaxation Dynamics}} of a {{Red Blood Cell}}}.
\bjtitle{Soft Matter}
\bvolume{17},
\bfpage{5978}--\blpage{5990}
(\byear{2021})
\doiurl{10.1039/D1SM00246E}
\end{barticle}
\endbibitem

\bibitem[\protect\citeauthoryear{{Puig-de-Morales-Marinkovic}
  et~al.}{2007}]{puig-de-morales-marinkovicViscoelasticityHumanRed2007b}
\begin{barticle}
\bauthor{\bsnm{{Puig-de-Morales-Marinkovic}}, \binits{M.}},
\bauthor{\bsnm{Turner}, \binits{K.T.}},
\bauthor{\bsnm{Butler}, \binits{J.P.}},
\bauthor{\bsnm{Fredberg}, \binits{J.J.}},
\bauthor{\bsnm{Suresh}, \binits{S.}}:
\batitle{Viscoelasticity of the {{Human Red Blood Cell}}}.
\bjtitle{American Journal of Physiology-Cell Physiology}
\bvolume{293}(\bissue{2}),
\bfpage{597}--\blpage{605}
(\byear{2007})
\doiurl{10.1152/ajpcell.00562.2006}
\end{barticle}
\endbibitem

\bibitem[\protect\citeauthoryear{Hochmuth
  et~al.}{1979}]{hochmuthRedCellExtensional1979}
\begin{barticle}
\bauthor{\bsnm{Hochmuth}, \binits{R.M.}},
\bauthor{\bsnm{Worthy}, \binits{P.R.}},
\bauthor{\bsnm{Evans}, \binits{E.A.}}:
\batitle{Red cell extensional recovery and the determination of membrane
  viscosity}.
\bjtitle{Biophysical Journal}
\bvolume{26}(\bissue{1}),
\bfpage{101}--\blpage{114}
(\byear{1979})
\doiurl{10.1016/S0006-3495(79)85238-8}
\end{barticle}
\endbibitem

\bibitem[\protect\citeauthoryear{Katchalsky
  et~al.}{1960}]{katchalskyRheologicalConsiderationsHaemolysing1960}
\begin{bchapter}
\bauthor{\bsnm{Katchalsky}, \binits{A.}},
\bauthor{\bsnm{Kedem}, \binits{O.}},
\bauthor{\bsnm{Klibansky}, \binits{C.}},
\bauthor{\bsnm{De~Vries}, \binits{A.}}:
\bctitle{Rheological considerations of the haemolysing red blood cell}.
In: \bbtitle{Flow Properties of Blood and Other Biological Systems},
pp. \bfpage{155}--\blpage{171}.
\bpublisher{Pergamon Press Inc.},
\blocation{New York}
(\byear{1960})
\end{bchapter}
\endbibitem

\bibitem[\protect\citeauthoryear{Rand}{1964}]{randMechanicalPropertiesRed1964}
\begin{barticle}
\bauthor{\bsnm{Rand}, \binits{R.P.}}:
\batitle{Mechanical {{Properties}} of the {{Red Cell Membrane}}: {{II}}.
  {{Viscoelastic Breakdown}} of the {{Membrane}}}.
\bjtitle{Biophysical Journal}
\bvolume{4}(\bissue{4}),
\bfpage{303}--\blpage{316}
(\byear{1964})
\doiurl{10.1016/S0006-3495(64)86784-9}
\end{barticle}
\endbibitem

\bibitem[\protect\citeauthoryear{Lommel
  et~al.}{}]{lommelExperimentalInvestigationApplicability}
\begin{botherref}
\oauthor{\bsnm{Lommel}, \binits{M.}},
\oauthor{\bsnm{Froese}, \binits{V.}},
\oauthor{\bsnm{Wolff}, \binits{H.}},
\oauthor{\bsnm{Dirkes}, \binits{N.}},
\oauthor{\bsnm{Vellguth}, \binits{K.}},
\oauthor{\bsnm{Behr}, \binits{M.}},
\oauthor{\bsnm{Kertzscher}, \binits{U.}}:
Experimental {{Investigation}} of the {{Applicability}} of the {{Stress-Based}}
  and {{Strain-Based Hemolysis Models}} for {{Short-Term Stress Peaks Typical}}
  for {{Rotary Blood Pumps}}.
Artificial Organs
\textbf{49}(7),
1108--1118
\doiurl{10.1111/aor.15002}
\end{botherref}
\endbibitem

\bibitem[\protect\citeauthoryear{Arwatz and
  Smits}{2013}]{arwatzViscoelasticModelShearInduced2013c}
\begin{barticle}
\bauthor{\bsnm{Arwatz}, \binits{G.}},
\bauthor{\bsnm{Smits}, \binits{A.J.}}:
\batitle{A {{Viscoelastic Model}} of {{Shear-Induced Hemolysis}} in {{Laminar
  Flow}}}.
\bjtitle{Biorheology}
\bvolume{50}(\bissue{1-2}),
\bfpage{45}--\blpage{55}
(\byear{2013})
\doiurl{10.3233/BIR-130626}
\end{barticle}
\endbibitem

\bibitem[\protect\citeauthoryear{Chen and
  Sharp}{2011}]{chenStrainBasedFlowInducedHemolysis2011a}
\begin{barticle}
\bauthor{\bsnm{Chen}, \binits{Y.}},
\bauthor{\bsnm{Sharp}, \binits{M.K.}}:
\batitle{A {{Strain-Based Flow-Induced Hemolysis Prediction Model Calibrated}}
  by {{In Vitro Erythrocyte Deformation Measurements}}}.
\bjtitle{Artificial Organs}
\bvolume{35}(\bissue{2}),
\bfpage{145}--\blpage{156}
(\byear{2011})
\doiurl{10.1111/j.1525-1594.2010.01050.x}
\end{barticle}
\endbibitem

\bibitem[\protect\citeauthoryear{Yeleswarapu
  et~al.}{1995}]{yeleswarapuMathematicalModelShearInduced1995}
\begin{barticle}
\bauthor{\bsnm{Yeleswarapu}, \binits{K.K.}},
\bauthor{\bsnm{Antaki}, \binits{J.F.}},
\bauthor{\bsnm{Kameneva}, \binits{M.V.}},
\bauthor{\bsnm{Rajagopal}, \binits{K.R.}}:
\batitle{A {{Mathematical Model}} for {{Shear-Induced Hemolysis}}}.
\bjtitle{Artificial Organs}
\bvolume{19}(\bissue{7}),
\bfpage{576}--\blpage{582}
(\byear{1995})
\doiurl{10.1111/j.1525-1594.1995.tb02384.x}
\end{barticle}
\endbibitem

\bibitem[\protect\citeauthoryear{Chen
  et~al.}{2013}]{chenTestingModelsFlowInduced2013c}
\begin{barticle}
\bauthor{\bsnm{Chen}, \binits{Y.}},
\bauthor{\bsnm{Kent}, \binits{T.L.}},
\bauthor{\bsnm{Sharp}, \binits{M.K.}}:
\batitle{Testing of {{Models}} of {{Flow-Induced Hemolysis}} in {{Blood Flow
  Through Hypodermic Needles}}}.
\bjtitle{Artificial Organs}
\bvolume{37}(\bissue{3}),
\bfpage{256}--\blpage{266}
(\byear{2013})
\doiurl{10.1111/j.1525-1594.2012.01569.x}
\end{barticle}
\endbibitem

\bibitem[\protect\citeauthoryear{Lee
  et~al.}{2019}]{leeEvaluationExtendedViscoelastic2019}
\begin{barticle}
\bauthor{\bsnm{Lee}, \binits{S.}},
\bauthor{\bsnm{Cho}, \binits{Y.}},
\bauthor{\bsnm{Kang}, \binits{S.}},
\bauthor{\bsnm{Hur}, \binits{N.}},
\bauthor{\bsnm{Kim}, \binits{W.}}:
\batitle{Evaluation of an extended viscoelastic model to predict hemolysis in
  cannulas and blood pumps}.
\bjtitle{Journal of Mechanical Science and Technology}
\bvolume{33}(\bissue{5}),
\bfpage{2181}--\blpage{2188}
(\byear{2019})
\doiurl{10.1007/s12206-019-0420-0}
\end{barticle}
\endbibitem

\bibitem[\protect\citeauthoryear{Yu
  et~al.}{2017}]{yuReviewHemolysisPrediction2017c}
\begin{barticle}
\bauthor{\bsnm{Yu}, \binits{H.}},
\bauthor{\bsnm{Engel}, \binits{S.}},
\bauthor{\bsnm{Janiga}, \binits{G.}},
\bauthor{\bsnm{Th{\'e}venin}, \binits{D.}}:
\batitle{A {{Review}} of {{Hemolysis Prediction Models}} for {{Computational
  Fluid Dynamics}}}.
\bjtitle{Artificial Organs}
\bvolume{41}(\bissue{7}),
\bfpage{603}--\blpage{621}
(\byear{2017})
\doiurl{10.1111/aor.12871}
\end{barticle}
\endbibitem

\bibitem[\protect\citeauthoryear{Dirkes
  et~al.}{2024}]{dirkesEulerianFormulationTensorbased2024}
\begin{barticle}
\bauthor{\bsnm{Dirkes}, \binits{N.}},
\bauthor{\bsnm{Key}, \binits{F.}},
\bauthor{\bsnm{Behr}, \binits{M.}}:
\batitle{Eulerian formulation of the tensor-based morphology equations for
  strain-based blood damage modeling}.
\bjtitle{Computer Methods in Applied Mechanics and Engineering}
\bvolume{426},
\bfpage{116979}
(\byear{2024})
\doiurl{10.1016/j.cma.2024.116979}
\end{barticle}
\endbibitem

\bibitem[\protect\citeauthoryear{Dirkes and
  Behr}{2025}]{dirkesSignificanceFlowVorticity2025}
\begin{bchapter}
\bauthor{\bsnm{Dirkes}, \binits{N.}},
\bauthor{\bsnm{Behr}, \binits{M.}}:
\bctitle{On the significance of flow vorticity for hemolysis modeling}.
In: \bbtitle{Topical Problems of Fluid Mechanics},
pp. \bfpage{51}--\blpage{58}.
\bpublisher{Institute of Thermomechanics of the Czech Academy of Sciences; CTU
  in Prague Faculty of Mech. Engineering Dept. Tech. Mathematics},
\blocation{Prague, Czechia}
(\byear{2025}).
\doiurl{10.14311/TPFM.2025.008}
\end{bchapter}
\endbibitem

\bibitem[\protect\citeauthoryear{Faghih and
  Sharp}{2020}]{faghihDeformationHumanRed2020b}
\begin{barticle}
\bauthor{\bsnm{Faghih}, \binits{M.M.}},
\bauthor{\bsnm{Sharp}, \binits{M.K.}}:
\batitle{Deformation of {{Human Red Blood Cells}} in {{Extensional Flow}}
  through a {{Hyperbolic Contraction}}}.
\bjtitle{Biomechanics and Modeling in Mechanobiology}
\bvolume{19}(\bissue{1}),
\bfpage{251}--\blpage{261}
(\byear{2020})
\doiurl{10.1007/s10237-019-01208-3}
\end{barticle}
\endbibitem

\bibitem[\protect\citeauthoryear{Faghih and
  Sharp}{2019}]{faghihModelingPredictionFlowinduced2019a}
\begin{barticle}
\bauthor{\bsnm{Faghih}, \binits{M.M.}},
\bauthor{\bsnm{Sharp}, \binits{M.K.}}:
\batitle{Modeling and prediction of flow-induced hemolysis: A review}.
\bjtitle{Biomechanics and Modeling in Mechanobiology}
\bvolume{18}(\bissue{4}),
\bfpage{845}--\blpage{881}
(\byear{2019})
\doiurl{10.1007/s10237-019-01137-1}
\end{barticle}
\endbibitem

\bibitem[\protect\citeauthoryear{Poorkhalil
  et~al.}{2016}]{poorkhalilNewApproachSemiempirical2016}
\begin{barticle}
\bauthor{\bsnm{Poorkhalil}, \binits{A.}},
\bauthor{\bsnm{Amoabediny}, \binits{G.}},
\bauthor{\bsnm{Tabesh}, \binits{H.}},
\bauthor{\bsnm{Behbahani}, \binits{M.}},
\bauthor{\bsnm{Mottaghy}, \binits{K.}}:
\batitle{A {{New Approach}} for {{Semiempirical Modeling}} of {{Mechanical
  Blood Trauma}}}.
\bjtitle{The International Journal of Artificial Organs}
\bvolume{39}(\bissue{4}),
\bfpage{171}--\blpage{177}
(\byear{2016})
\doiurl{10.5301/ijao.5000474}
\end{barticle}
\endbibitem

\bibitem[\protect\citeauthoryear{McKean}{2020}]{mckeanDevelopmentHemolysisModel2020}
\begin{botherref}
\oauthor{\bsnm{McKean}, \binits{A.}}:
Development of a {{Hemolysis Model}} with {{Sublethal Hemoglobin Release}}.
Master's thesis,
McGill University
(2020)
\end{botherref}
\endbibitem

\bibitem[\protect\citeauthoryear{Giersiepen
  et~al.}{1990}]{giersiepenEstimationShearStressRelated1990a}
\begin{barticle}
\bauthor{\bsnm{Giersiepen}, \binits{M.}},
\bauthor{\bsnm{Wurzinger}, \binits{L.J.}},
\bauthor{\bsnm{Opitz}, \binits{R.}},
\bauthor{\bsnm{Reul}, \binits{H.}}:
\batitle{Estimation of {{Shear Stress-Related Blood Damage}} in {{Heart Valve
  Prostheses}}---{{In Vitro Comparison}} of 25 {{Aortic Valves}}}.
\bjtitle{International Journal of Artificial Organs}
\bvolume{13}(\bissue{5}),
\bfpage{300}--\blpage{306}
(\byear{1990})
\doiurl{10.1177/039139889001300507}
\end{barticle}
\endbibitem

\bibitem[\protect\citeauthoryear{Ding
  et~al.}{2015}]{dingShearInducedHemolysisSpecies2015a}
\begin{barticle}
\bauthor{\bsnm{Ding}, \binits{J.}},
\bauthor{\bsnm{Niu}, \binits{S.}},
\bauthor{\bsnm{Chen}, \binits{Z.}},
\bauthor{\bsnm{Zhang}, \binits{T.}},
\bauthor{\bsnm{Griffith}, \binits{B.P.}},
\bauthor{\bsnm{Wu}, \binits{Z.J.}}:
\batitle{Shear-{{Induced Hemolysis}}: {{Species Differences}}}.
\bjtitle{Artificial Organs}
\bvolume{39}(\bissue{9}),
\bfpage{795}--\blpage{802}
(\byear{2015})
\doiurl{10.1111/aor.12459}
\end{barticle}
\endbibitem

\bibitem[\protect\citeauthoryear{Vitale
  et~al.}{2014}]{vitaleMultiscaleBiophysicalModel2014a}
\begin{barticle}
\bauthor{\bsnm{Vitale}, \binits{F.}},
\bauthor{\bsnm{Nam}, \binits{J.}},
\bauthor{\bsnm{Turchetti}, \binits{L.}},
\bauthor{\bsnm{Behr}, \binits{M.}},
\bauthor{\bsnm{Raphael}, \binits{R.}},
\bauthor{\bsnm{Annesini}, \binits{M.C.}},
\bauthor{\bsnm{Pasquali}, \binits{M.}}:
\batitle{A multiscale, biophysical model of flow-induced red blood cell
  damage}.
\bjtitle{AIChE Journal}
\bvolume{60}(\bissue{4}),
\bfpage{1509}--\blpage{1516}
(\byear{2014})
\doiurl{10.1002/aic.14318}
\end{barticle}
\endbibitem

\bibitem[\protect\citeauthoryear{Sohrabi and
  Liu}{2017}]{sohrabiCellularModelShearInduced2017b}
\begin{barticle}
\bauthor{\bsnm{Sohrabi}, \binits{S.}},
\bauthor{\bsnm{Liu}, \binits{Y.}}:
\batitle{A {{Cellular Model}} of {{Shear-Induced Hemolysis}}}.
\bjtitle{Artificial Organs}
\bvolume{41}(\bissue{9}),
\bfpage{80}--\blpage{91}
(\byear{2017})
\doiurl{10.1111/aor.12832}
\end{barticle}
\endbibitem

\bibitem[\protect\citeauthoryear{Ha{\ss}ler
  et~al.}{2020}]{hasslerFiniteelementFormulationAdvection2020a}
\begin{barticle}
\bauthor{\bsnm{Ha{\ss}ler}, \binits{S.}},
\bauthor{\bsnm{Ranno}, \binits{A.M.}},
\bauthor{\bsnm{Behr}, \binits{M.}}:
\batitle{Finite-element formulation for advection--reaction equations with
  change of variable and discontinuity capturing}.
\bjtitle{Computer Methods in Applied Mechanics and Engineering}
\bvolume{369},
\bfpage{113171}
(\byear{2020})
\doiurl{10.1016/j.cma.2020.113171}
\end{barticle}
\endbibitem

\bibitem[\protect\citeauthoryear{Pauli}{2016}]{pauliStabilizedFiniteElement2016c}
\begin{botherref}
\oauthor{\bsnm{Pauli}, \binits{L.}}:
Stabilized {{Finite Element Methods}} for {{Computational Design}} of
  {{Blood-Handling Devices}}.
PhD thesis,
RWTH Aachen University,
Aachen, Germany
(2016)
\end{botherref}
\endbibitem

\bibitem[\protect\citeauthoryear{Melka
  et~al.}{2019}]{melkaNumericalInvestigationMultiphase2019}
\begin{barticle}
\bauthor{\bsnm{Melka}, \binits{B.}},
\bauthor{\bsnm{Adamczyk}, \binits{W.P.}},
\bauthor{\bsnm{Rojczyk}, \binits{M.}},
\bauthor{\bsnm{Nowak}, \binits{M.L.}},
\bauthor{\bsnm{Gracka}, \binits{M.}},
\bauthor{\bsnm{Nowak}, \binits{A.J.}},
\bauthor{\bsnm{Golda}, \binits{A.}},
\bauthor{\bsnm{Bialecki}, \binits{R.A.}},
\bauthor{\bsnm{Ostrowski}, \binits{Z.}}:
\batitle{Numerical investigation of multiphase blood flow coupled with lumped
  parameter model of outflow}.
\bjtitle{International Journal of Numerical Methods for Heat \& Fluid Flow}
\bvolume{30}(\bissue{1}),
\bfpage{228}--\blpage{244}
(\byear{2019})
\doiurl{10.1108/HFF-04-2019-0279}
\end{barticle}
\endbibitem

\bibitem[\protect\citeauthoryear{Bodn{\'a}r
  et~al.}{2011}]{bodnarSimulationThreeDimensionalFlow2011}
\begin{barticle}
\bauthor{\bsnm{Bodn{\'a}r}, \binits{T.}},
\bauthor{\bsnm{Rajagopal}, \binits{K.R.}},
\bauthor{\bsnm{Sequeira}, \binits{A.}}:
\batitle{Simulation of the {{Three-Dimensional Flow}} of {{Blood Using}} a
  {{Shear-Thinning Viscoelastic Fluid Model}}}.
\bjtitle{Mathematical Modelling of Natural Phenomena}
\bvolume{6}(\bissue{5}),
\bfpage{1}--\blpage{24}
(\byear{2011})
\doiurl{10.1051/mmnp/20116501}
\end{barticle}
\endbibitem

\bibitem[\protect\citeauthoryear{Krisher
  et~al.}{2022}]{krisherEffectBloodViscosity2022}
\begin{barticle}
\bauthor{\bsnm{Krisher}, \binits{J.A.}},
\bauthor{\bsnm{Malinauskas}, \binits{R.A.}},
\bauthor{\bsnm{Day}, \binits{S.W.}}:
\batitle{The effect of blood viscosity on shear-induced hemolysis using a
  magnetically levitated shearing device}.
\bjtitle{Artificial Organs}
\bvolume{46}(\bissue{6}),
\bfpage{1027}--\blpage{1039}
(\byear{2022})
\doiurl{10.1111/aor.14172}
\end{barticle}
\endbibitem

\bibitem[\protect\citeauthoryear{Bludszuweit}{1995}]{bludszuweitModelGeneralMechanical1995c}
\begin{barticle}
\bauthor{\bsnm{Bludszuweit}, \binits{C.}}:
\batitle{Model for a {{General Mechanical Blood Damage Prediction}}}.
\bjtitle{Artificial Organs}
\bvolume{19}(\bissue{7}),
\bfpage{583}--\blpage{589}
(\byear{1995})
\doiurl{10.1111/j.1525-1594.1995.tb02385.x}
\end{barticle}
\endbibitem

\bibitem[\protect\citeauthoryear{H{\'e}non
  et~al.}{1999}]{henonNewDeterminationShear1999}
\begin{barticle}
\bauthor{\bsnm{H{\'e}non}, \binits{S.}},
\bauthor{\bsnm{Lenormand}, \binits{G.}},
\bauthor{\bsnm{Richert}, \binits{A.}},
\bauthor{\bsnm{Gallet}, \binits{F.}}:
\batitle{A {{New Determination}} of the {{Shear Modulus}} of the {{Human
  Erythrocyte Membrane Using Optical Tweezers}}}.
\bjtitle{Biophysical Journal}
\bvolume{76}(\bissue{2}),
\bfpage{1145}--\blpage{1151}
(\byear{1999})
\doiurl{10.1016/S0006-3495(99)77279-6}
\end{barticle}
\endbibitem

\bibitem[\protect\citeauthoryear{Bronkhorst
  et~al.}{1995}]{bronkhorstNewMethodStudy1995}
\begin{barticle}
\bauthor{\bsnm{Bronkhorst}, \binits{P.J.}},
\bauthor{\bsnm{Streekstra}, \binits{G.J.}},
\bauthor{\bsnm{Grimbergen}, \binits{J.}},
\bauthor{\bsnm{Nijhof}, \binits{E.J.}},
\bauthor{\bsnm{Sixma}, \binits{J.J.}},
\bauthor{\bsnm{Brakenhoff}, \binits{G.J.}}:
\batitle{A new method to study shape recovery of red blood cells using multiple
  optical trapping}.
\bjtitle{Biophysical Journal}
\bvolume{69}(\bissue{5}),
\bfpage{1666}--\blpage{1673}
(\byear{1995})
\doiurl{10.1016/S0006-3495(95)80084-6}
\end{barticle}
\endbibitem

\bibitem[\protect\citeauthoryear{Arora
  et~al.}{2004}]{aroraTensorBasedMeasureEstimating2004}
\begin{barticle}
\bauthor{\bsnm{Arora}, \binits{D.}},
\bauthor{\bsnm{Behr}, \binits{M.}},
\bauthor{\bsnm{Pasquali}, \binits{M.}}:
\batitle{A {{Tensor-Based Measure}} for {{Estimating Blood Damage}}}.
\bjtitle{Artificial Organs}
\bvolume{28}(\bissue{11}),
\bfpage{1002}--\blpage{1015}
(\byear{2004})
\doiurl{10.1111/j.1525-1594.2004.00072.x}
\end{barticle}
\endbibitem

\bibitem[\protect\citeauthoryear{Ha{\ss}ler
  et~al.}{2019}]{hasslerVariationalMultiscaleFormulation2019b}
\begin{barticle}
\bauthor{\bsnm{Ha{\ss}ler}, \binits{S.}},
\bauthor{\bsnm{Pauli}, \binits{L.}},
\bauthor{\bsnm{Behr}, \binits{M.}}:
\batitle{The {{Variational Multiscale Formulation}} for the {{Fully-Implicit
  Log-Morphology Equation}} as a {{Tensor-Based Blood Damage Model}}}.
\bjtitle{International Journal for Numerical Methods in Biomedical Engineering}
\bvolume{35}(\bissue{12}),
\bfpage{3262}
(\byear{2019})
\doiurl{10.1002/cnm.3262}
\end{barticle}
\endbibitem

\bibitem[\protect\citeauthoryear{Taskin
  et~al.}{2012}]{taskinEvaluationEulerianLagrangian2012b}
\begin{barticle}
\bauthor{\bsnm{Taskin}, \binits{M.E.}},
\bauthor{\bsnm{Fraser}, \binits{K.H.}},
\bauthor{\bsnm{Zhang}, \binits{T.}},
\bauthor{\bsnm{Wu}, \binits{C.}},
\bauthor{\bsnm{Griffith}, \binits{B.P.}},
\bauthor{\bsnm{Wu}, \binits{Z.J.}}:
\batitle{Evaluation of {{Eulerian}} and {{Lagrangian}} models for hemolysis
  estimation}.
\bjtitle{ASAIO journal (American Society for Artificial Internal Organs)}
\bvolume{58}(\bissue{4}),
\bfpage{363}--\blpage{372}
(\byear{2012})
\doiurl{10.1097/MAT.0b013e318254833b}
\end{barticle}
\endbibitem

\bibitem[\protect\citeauthoryear{Blum
  et~al.}{2025}]{blumQuantifyingExperimentalVariability2025}
\begin{barticle}
\bauthor{\bsnm{Blum}, \binits{C.}},
\bauthor{\bsnm{Mous}, \binits{M.}},
\bauthor{\bsnm{Steinseifer}, \binits{U.}},
\bauthor{\bsnm{Clauser}, \binits{J.C.}},
\bauthor{\bsnm{Neidlin}, \binits{M.}}:
\batitle{Quantifying {{Experimental Variability}} in {{Shear-Induced
  Hemolysis}} to {{Support Uncertainty-Aware Hemolysis Models}}}.
\bjtitle{Annals of Biomedical Engineering}
\bvolume{53}(\bissue{10}),
\bfpage{2551}--\blpage{2561}
(\byear{2025})
\doiurl{10.1007/s10439-025-03786-z}
\end{barticle}
\endbibitem

\bibitem[\protect\citeauthoryear{Faghih
  et~al.}{2023}]{faghihPracticalImplicationsErroneous2023}
\begin{barticle}
\bauthor{\bsnm{Faghih}, \binits{M.M.}},
\bauthor{\bsnm{Craven}, \binits{B.A.}},
\bauthor{\bsnm{Sharp}, \binits{M.K.}}:
\batitle{Practical implications of the erroneous treatment of exposure time in
  the {{Eulerian}} hemolysis power law model}.
\bjtitle{Artificial Organs}
\bvolume{47}(\bissue{9}),
\bfpage{1531}--\blpage{1538}
(\byear{2023})
\doiurl{10.1111/aor.14543}
\end{barticle}
\endbibitem

\bibitem[\protect\citeauthoryear{Lacasse
  et~al.}{2007}]{lacasseMechanicalHemolysisBlood2007}
\begin{barticle}
\bauthor{\bsnm{Lacasse}, \binits{D.}},
\bauthor{\bsnm{Garon}, \binits{A.}},
\bauthor{\bsnm{Pelletier}, \binits{D.}}:
\batitle{Mechanical hemolysis in blood flow: User-independent predictions with
  the solution of a partial differential equation}.
\bjtitle{Computer Methods in Biomechanics and Biomedical Engineering}
\bvolume{10}(\bissue{1}),
\bfpage{1}--\blpage{12}
(\byear{2007})
\doiurl{10.1080/10255840600985535}
\end{barticle}
\endbibitem

\bibitem[\protect\citeauthoryear{Craven
  et~al.}{2019}]{cravenCFDbasedKrigingSurrogate2019}
\begin{barticle}
\bauthor{\bsnm{Craven}, \binits{B.A.}},
\bauthor{\bsnm{Aycock}, \binits{K.I.}},
\bauthor{\bsnm{Herbertson}, \binits{L.H.}},
\bauthor{\bsnm{Malinauskas}, \binits{R.A.}}:
\batitle{A {{CFD-based Kriging}} surrogate modeling approach for predicting
  device-specific hemolysis power law coefficients in blood-contacting medical
  devices}.
\bjtitle{Biomechanics and Modeling in Mechanobiology}
\bvolume{18}(\bissue{4}),
\bfpage{1005}--\blpage{1030}
(\byear{2019})
\doiurl{10.1007/s10237-019-01126-4}
\end{barticle}
\endbibitem

\bibitem[\protect\citeauthoryear{Garon and
  Farinas}{2004}]{garonFastThreedimensionalNumerical2004a}
\begin{barticle}
\bauthor{\bsnm{Garon}, \binits{A.}},
\bauthor{\bsnm{Farinas}, \binits{M.-I.}}:
\batitle{Fast three-dimensional numerical hemolysis approximation}.
\bjtitle{Artificial Organs}
\bvolume{28}(\bissue{11}),
\bfpage{1016}--\blpage{1025}
(\byear{2004})
\doiurl{10.1111/j.1525-1594.2004.00026.x}
\end{barticle}
\endbibitem

\bibitem[\protect\citeauthoryear{Pauli
  et~al.}{2015}]{pauliStabilizedFiniteElement2015b}
\begin{barticle}
\bauthor{\bsnm{Pauli}, \binits{L.}},
\bauthor{\bsnm{Both}, \binits{J.}},
\bauthor{\bsnm{Behr}, \binits{M.}}:
\batitle{Stabilized {{Finite Element Method}} for {{Flows}} with {{Multiple
  Reference Frames}}}.
\bjtitle{International Journal for Numerical Methods in Fluids}
\bvolume{78}(\bissue{11}),
\bfpage{657}--\blpage{669}
(\byear{2015})
\doiurl{10.1002/fld.4032}
\end{barticle}
\endbibitem

\bibitem[\protect\citeauthoryear{Bazilevs
  et~al.}{2007}]{bazilevsVariationalMultiscaleResidualbased2007a}
\begin{barticle}
\bauthor{\bsnm{Bazilevs}, \binits{Y.}},
\bauthor{\bsnm{Calo}, \binits{V.M.}},
\bauthor{\bsnm{Cottrell}, \binits{J.A.}},
\bauthor{\bsnm{Hughes}, \binits{T.J.R.}},
\bauthor{\bsnm{Reali}, \binits{A.}},
\bauthor{\bsnm{Scovazzi}, \binits{G.}}:
\batitle{Variational multiscale residual-based turbulence modeling for large
  eddy simulation of incompressible flows}.
\bjtitle{Computer Methods in Applied Mechanics and Engineering}
\bvolume{197}(\bissue{1}),
\bfpage{173}--\blpage{201}
(\byear{2007})
\doiurl{10.1016/j.cma.2007.07.016}
\end{barticle}
\endbibitem

\bibitem[\protect\citeauthoryear{Malinauskas
  et~al.}{}]{malinauskasFDABenchmarkMedical}
\begin{botherref}
\oauthor{\bsnm{Malinauskas}, \binits{R.A.}},
\oauthor{\bsnm{Hariharan}, \binits{P.}},
\oauthor{\bsnm{Day}, \binits{S.W.}},
\oauthor{\bsnm{Herbertson}, \binits{L.H.}},
\oauthor{\bsnm{Buesen}, \binits{M.}},
\oauthor{\bsnm{Steinseifer}, \binits{U.}},
\oauthor{\bsnm{Aycock}, \binits{K.I.}},
\oauthor{\bsnm{Good}, \binits{B.C.}},
\oauthor{\bsnm{Deutsch}, \binits{S.}},
\oauthor{\bsnm{Manning}, \binits{K.B.}},
\oauthor{\bsnm{Craven}, \binits{B.A.}}:
{{FDA Benchmark Medical Device Flow Models}} for {{CFD Validation}}.
ASAIO Journal
\textbf{63}(2),
150
\doiurl{10.1097/MAT.0000000000000499}
\end{botherref}
\endbibitem

\bibitem[\protect\citeauthoryear{Hariharan
  et~al.}{2011}]{hariharanMultilaboratoryParticleImage2011a}
\begin{botherref}
\oauthor{\bsnm{Hariharan}, \binits{P.}},
\oauthor{\bsnm{Giarra}, \binits{M.}},
\oauthor{\bsnm{Reddy}, \binits{V.}},
\oauthor{\bsnm{Day}, \binits{S.W.}},
\oauthor{\bsnm{Manning}, \binits{K.B.}},
\oauthor{\bsnm{Deutsch}, \binits{S.}},
\oauthor{\bsnm{Stewart}, \binits{S.F.C.}},
\oauthor{\bsnm{Myers}, \binits{M.R.}},
\oauthor{\bsnm{Berman}, \binits{M.R.}},
\oauthor{\bsnm{Burgreen}, \binits{G.W.}},
\oauthor{\bsnm{Paterson}, \binits{E.G.}},
\oauthor{\bsnm{Malinauskas}, \binits{R.A.}}:
Multilaboratory {{Particle Image Velocimetry Analysis}} of the {{FDA Benchmark
  Nozzle Model}} to {{Support Validation}} of {{Computational Fluid Dynamics
  Simulations}}.
Journal of Biomechanical Engineering
\textbf{133}(041002)
(2011)
\doiurl{10.1115/1.4003440}
\end{botherref}
\endbibitem

\bibitem[\protect\citeauthoryear{Mantegazza
  et~al.}{2023}]{mantegazzaExaminingUniversalityHemolysis2023b}
\begin{barticle}
\bauthor{\bsnm{Mantegazza}, \binits{A.}},
\bauthor{\bsnm{Tobin}, \binits{N.}},
\bauthor{\bsnm{Manning}, \binits{K.B.}},
\bauthor{\bsnm{Craven}, \binits{B.A.}}:
\batitle{Examining the universality of the hemolysis power law model from
  simulations of the {{FDA}} nozzle using calibrated model coefficients}.
\bjtitle{Biomechanics and Modeling in Mechanobiology}
\bvolume{22}(\bissue{2}),
\bfpage{433}--\blpage{451}
(\byear{2023})
\doiurl{10.1007/s10237-022-01655-5}
\end{barticle}
\endbibitem

\bibitem[\protect\citeauthoryear{Herbertson
  et~al.}{2015}]{herbertsonMultilaboratoryStudyFlowInduced2015}
\begin{barticle}
\bauthor{\bsnm{Herbertson}, \binits{L.H.}},
\bauthor{\bsnm{Olia}, \binits{S.E.}},
\bauthor{\bsnm{Daly}, \binits{A.}},
\bauthor{\bsnm{Noatch}, \binits{C.P.}},
\bauthor{\bsnm{Smith}, \binits{W.A.}},
\bauthor{\bsnm{Kameneva}, \binits{M.V.}},
\bauthor{\bsnm{Malinauskas}, \binits{R.A.}}:
\batitle{Multilaboratory {{Study}} of {{Flow-Induced Hemolysis Using}} the
  {{FDA Benchmark Nozzle Model}}}.
\bjtitle{Artificial Organs}
\bvolume{39}(\bissue{3}),
\bfpage{237}--\blpage{248}
(\byear{2015})
\doiurl{10.1111/aor.12368}
\end{barticle}
\endbibitem

\bibitem[\protect\citeauthoryear{Goubergrits
  et~al.}{2016}]{goubergritsTurbulenceBloodDamage2016}
\begin{barticle}
\bauthor{\bsnm{Goubergrits}, \binits{L.}},
\bauthor{\bsnm{Osman}, \binits{J.}},
\bauthor{\bsnm{Mevert}, \binits{R.}},
\bauthor{\bsnm{Kertzscher}, \binits{U.}},
\bauthor{\bsnm{P{\"o}thkow}, \binits{K.}},
\bauthor{\bsnm{Hege}, \binits{H.-C.}}:
\batitle{Turbulence in {{Blood Damage Modeling}}}.
\bjtitle{The International Journal of Artificial Organs}
\bvolume{39}(\bissue{4}),
\bfpage{160}--\blpage{165}
(\byear{2016})
\doiurl{10.5301/ijao.5000476}
\end{barticle}
\endbibitem

\bibitem[\protect\citeauthoryear{Wu
  et~al.}{2019}]{wuRepresentationEffectiveStress2019}
\begin{barticle}
\bauthor{\bsnm{Wu}, \binits{P.}},
\bauthor{\bsnm{Gao}, \binits{Q.}},
\bauthor{\bsnm{Hsu}, \binits{P.-L.}}:
\batitle{On the representation of effective stress for computing hemolysis}.
\bjtitle{Biomechanics and Modeling in Mechanobiology}
\bvolume{18}(\bissue{3}),
\bfpage{665}--\blpage{679}
(\byear{2019})
\doiurl{10.1007/s10237-018-01108-y}
\end{barticle}
\endbibitem

\bibitem[\protect\citeauthoryear{Blum
  et~al.}{2025}]{blumUncertaintyAwareHemolysisModeling2025a}
\begin{barticle}
\bauthor{\bsnm{Blum}, \binits{C.}},
\bauthor{\bsnm{Steinseifer}, \binits{U.}},
\bauthor{\bsnm{Neidlin}, \binits{M.}}:
\batitle{Toward {{Uncertainty-Aware Hemolysis Modeling}}: {{A Universal
  Approach}} to {{Address Experimental Variance}}}.
\bjtitle{International Journal for Numerical Methods in Biomedical Engineering}
\bvolume{41}(\bissue{5}),
\bfpage{70040}
(\byear{2025})
\doiurl{10.1002/cnm.70040}
\end{barticle}
\endbibitem

\bibitem[\protect\citeauthoryear{Keller}{1979}]{kellerSurfaceAreaEllipsoid1979}
\begin{barticle}
\bauthor{\bsnm{Keller}, \binits{S.R.}}:
\batitle{On the {{Surface Area}} of the {{Ellipsoid}}}.
\bjtitle{Mathematics of Computation}
\bvolume{33}(\bissue{145}),
\bfpage{310}--\blpage{314}
(\byear{1979})
\doiurl{10.2307/2006043}
\end{barticle}
\endbibitem

\end{thebibliography}

\end{document}